\documentclass[
 reprint,
 amsmath,
 aps,
 prx,
nofootinbib,
floatfix
]{revtex4-2}

\usepackage{graphicx}%
\usepackage{dcolumn}%
\usepackage{bm}%
\usepackage[utf8]{inputenc} %
\usepackage[T1]{fontenc}    %
\usepackage{hyperref}       %
\usepackage{url}            %
\usepackage{booktabs}       %
\usepackage{amsfonts}       %
\usepackage{microtype}      %
\usepackage{amssymb,amsmath,amsthm,graphicx}
\usepackage{xcolor}
\usepackage{comment}
\usepackage{multirow}
\usepackage[caption=false]{subfig}
\usepackage{float}
\usepackage{tabularx}
\usepackage{float}
\usepackage{physics}
\makeatletter
\let\newfloat\newfloat@ltx
\makeatother
\usepackage{algorithm}
\usepackage{algpseudocode}
\usepackage{enumitem}

\algrenewcommand\algorithmicrequire{\textbf{Input:}}
\algrenewcommand\algorithmicensure{\textbf{Output:}}
\newcommand{\ReturnX}{\State\textbf{return:} }
\begin{document}

\title{
Integrated error-suppressed pipeline for quantum optimization of nontrivial binary combinatorial optimization problems on gate-model hardware at the 156-qubit scale
}

\author{Natasha Sachdeva}
\email{natasha.sachdeva@q-ctrl.com}
\author{Gavin S. Hartnett}
\author{Smarak Maity}
\author{Samuel Marsh}
\author{Yulun Wang}
\author{Adam Winick}
\author{Ryan Dougherty}
\author{Daniel Canuto}
\author{You Quan Chong}
\author{G. Adam Cox}
\author{Michael Hush}
\author{Pranav S. Mundada}
\author{Christopher D. B. Bentley}
\author{Michael J. Biercuk}
\author{Yuval Baum}
\affiliation{Q-CTRL, Sydney, NSW Australia, and Los Angeles, CA USA}

\begin{abstract}
We introduce a novel hybrid quantum-classical variational optimization method for unconstrained binary combinatorial optimization problems on gate-model quantum computers, integrating a custom variational ansatz, staged feedback-based dual variational parameter update strategies, efficient parametric compilation, automated error suppression during hardware execution, and scalable O($n$) classical post-processing to correct for bitflip errors. Without this integrated approach, we show that standard circuit execution at scale produces output indistinguishable from random sampling, establishing the necessity of each pipeline component. We benchmark the method on IBM superconducting quantum computers for several classically nontrivial optimization problems, where the optimization is conducted on hardware with no use of classical simulation or prior knowledge of the solution. For Max-Cut on random regular graphs with topologies not matched to device connectivity, the method achieves approximation ratios of 100\% for unweighted 3-regular graphs up to 156 nodes, weighted regular graphs up to 80 nodes, and weighted 7-regular graphs up 50 nodes. Applied to higher-order binary optimization, the method finds the ground state energy of 127- and 156-qubit spin-glass models matched to device topology with linear, quadratic, and cubic interaction terms, achieving approximation ratios of at least 99.5\% across all instances tested. The method consistently outperforms a classical local solver across all problems. Where published results on identical problem instances are available, our method demonstrates competitive or superior performance. These results demonstrate that an appropriately engineered variational approach enables gate-model quantum computers to produce high-quality solutions for nontrivial binary optimization problems at the 156 qubit scale, where na\"\i ve implementations are insufficient for good performance. 
\end{abstract}

\maketitle

\section{Introduction}

Quantum optimization continues to be a central motivation for quantum computing, with industry-relevant combinatorial optimization problems arising across logistics, finance, transportation, networking, and other industries \cite{Gao2017}. Although long-term progress toward fault-tolerant devices is expected to unlock the most rigorous quantum speedups, understanding how optimization algorithms perform on today’s noisy hardware remains an active and scientifically important area of research. Noisy intermediate-scale quantum (NISQ) era \cite{preskill2018nisq} algorithm development enables systematic exploration of hardware-efficient circuit structures, hybrid quantum-classical optimization strategies, and error-aware execution techniques, providing insight into both the opportunities and limitations of near-term devices. Such investigations help shape the design of future fault-tolerant algorithms, refine resource estimates, and identify algorithmic principles that remain valuable across hardware regimes. Consequently, pushing the scale and robustness of quantum optimization on existing quantum processors is a meaningful step toward practical quantum advantage and continues to motivate the development of new hybrid methodologies.

Many important optimization tasks remain classically challenging at moderate scales—typically hundreds of binary variables with dense interactions or higher-order terms. Widely used solvers such as CPLEX \cite{cplex} and Gurobi \cite{gurobi}, along with more recent specialized heuristics \cite{rehfeldt2023faster}, represent decades of algorithmic engineering and provide state-of-the-art performance for large classes of quadratic or linearly constrained optimization problems. Although these solvers are, in principle, exact and capable of certifying global optimality, achieving such guarantees can require substantial computational resources for combinatorially complex instances. As a result, they are frequently operated under practical time or memory limits, where the emphasis shifts from provable optimality to high-quality feasible solutions. Their effectiveness further diminishes when moving beyond quadratic formulations, when constraints induce strong coupling, or when higher-order interactions create rugged energy landscapes. For medium-scale instances on the order of $10^2$--$10^3$ binary variables, these factors can lead to solutions that remain far from optimal within acceptable run times, as observed in real-world applications such as periodic train scheduling where acceptable run-time can under-perform the true optimal solutions by up to 50\% \cite{BORNDORFER2020}. These limitations motivate continued exploration of alternative computational paradigms, including quantum approaches.

Quantum optimization provides a framework for representing both quadratic and higher-order binary problems as Ising Hamiltonians suitable for gate-model and annealing-based devices. A key objective is to develop algorithmic techniques that make effective use of today’s hardware while remaining scalable to larger processors as they become available. Although recent hardware roadmaps suggest a more gradual trajectory toward devices with several hundred to a thousand high-fidelity qubits, resource analyses indicate that systems in this regime may provide performance beyond that of optimized classical heuristics for certain problem classes \cite{boulebnane2022, VIDAL2014, Guerreschi2019, shaydulin2023evidence, BLEKOS2024}. Consequently, evaluating the behavior of quantum optimization algorithms on present-day processors offers valuable insight into their scalability and robustness.

Gate-model quantum computers can execute quadratic unconstrained binary optimization (QUBO) problems and facilitate direct incorporation of higher-order optimization terms and constraints \cite{Yarkoni_2022, deng2023,verchere2023, Glos2022, Pradeep2022, Fuchs_2022}. However, the generality of the hardware introduces numerous error pathways that significantly curtail useful system size and performance. The quantum approximate optimization algorithm (QAOA) \cite{farhi2014quantum} provides a natural variational framework for these problems, but achieving high-quality solutions on hardware has proven challenging. On superconducting qubit devices, QAOA applied to problems with non-hardware-native graph topologies has produced lackluster results at scales up to 40 qubits \cite{Niu2022, Weidenfeller2022scalingofquantum, Harrigan2020, Sack2024}, with standard implementations returning output that is often indistinguishable from random sampling. In trapped-ion processors, successful QAOA implementations capable of returning optimal results have reached approximately 32 qubits, although in Ref.~\cite{shaydulin2023qaoa}, this required classical simulation to pre-calculate optimized circuit parameters and relied on high-depth circuits ($p\geq 10$). A full on-hardware optimization of $p=2$ QAOA for 32-node Max-Cut returning the optimal cut was demonstrated in Ref.~\cite{moses2023racetrack}, and a qubit-reuse strategy on 20--32 qubit trapped-ion devices has extended QAOA demonstrations to 130-nodes, although without finding optimal solutions %
\cite{decross2023qubitreuse, moses2023racetrack}.

More recently, several works have expanded the landscape of quantum optimization. In Ref.~\cite{shaydulin2024_labs} performed noiseless simulations of QAOA on the low-autocorrelation binary sequences (LABS) problem with up to 40 qubits, finding that QAOA runtime scaled better than branch-and-bound solvers, and demonstrated experimental progress in executing the algorithm on Quantinuum trapped-ion processors using algorithm-specific error detection. In Ref.~\cite{Sciorilli2025}, the authors introduced a qubit-efficient encoding that maps O($n^k$) binary variables onto $n$ qubits, achieving approximation ratios estimated to exceed the computational hardness threshold of 0.941 for Max-Cut instances with $m=2000$ variables on 17 trapped-ion qubits. In Ref.~\cite{montanezbarrera2024}, the authors showed that fixed linear-ramp QAOA schedules achieve competitive performance without per-instance parameter optimization, demonstrating the protocol on multiple QPU platforms with up to 109 qubits and circuits requiring over 21,000 CNOT gates. In a separate cross-platform benchmarking study \cite{montanezbarrera2025benchmarking}, QAOA-based benchmarks were evaluated across 19 QPUs from five vendors, finding that Quantinuum H2-1 maintained coherent computation on a fully connected 56-qubit Max-Cut instance. These advances collectively illustrate both the rapid evolution of the field and the diversity of approaches from qubit-efficient encodings that maximize problem size per physical qubit, to methods that eliminate the classical optimization loop, to brute-force depth scaling on hardware with sufficient coherence. 

Quantum annealing (QA) provides an alternative approach to binary optimization \cite{Yarkoni_2022, Hauke_2020} in which the limited functionality of the hardware reduces error pathways and simplifies device scaling. While gate-model quantum computers and quantum annealers share formal equivalence from the perspective of computational complexity \cite{Aharonov2007}, practical comparisons have generally favored annealers for quadratic unconstrained binary optimization (QUBO) problems~\cite{Zbinden2020, Lopez-Bezanilla2024}. Higher-order terms and/or constraints have also been included by leveraging QA scaling advantages via the use of slack variables and penalty terms \cite{junger2021quantum, tasseff2022emerging, pelofske2023quantum}. In direct comparisons between standard QAOA implementations on gate-model hardware and quantum annealers, annealers have typically delivered superior solutions~\cite{pelofske2024short, lubinski2024, mohseni2024}. In Ref.~\cite{pelofske2024short}, the authors confirm this pattern at the 127-qubit scale on $ibm\_washington$, finding that short-depth QAOA ($p=$1,2) was outperformed by D-Wave quantum annealers across all higher-order Ising model instances examined, which were problems where the connectivity matched the gate-model device's native heavy-hex topology. A central question, then, is whether this performance gap reflects a fundamental limitation of gate-model optimization or whether it can be bridged through more careful integration of algorithmic design with hardware error management. We investigate these problem instances, along with Max-Cut on random regular graphs, in this work.

\begin{figure*}[t]
    \centering
    \includegraphics[width=1.0\textwidth]{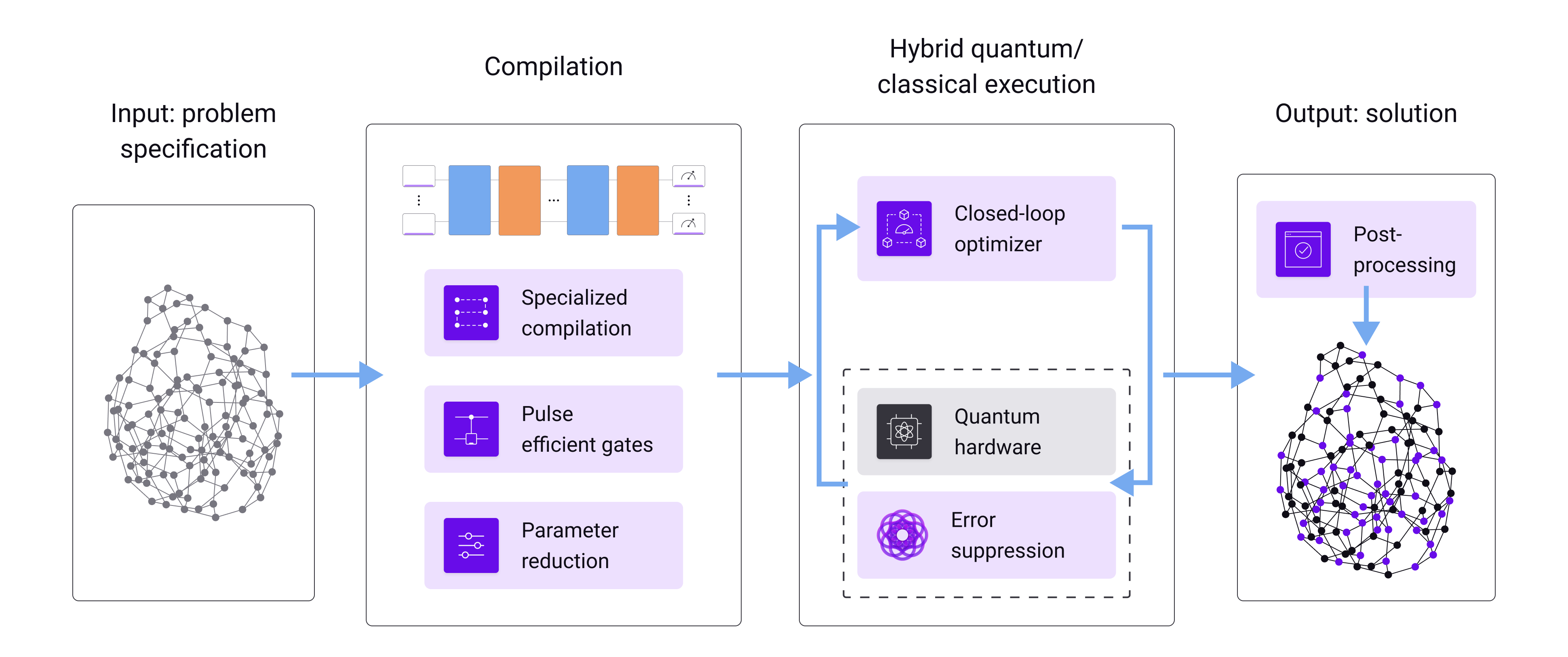}
    \caption{
    \label{fig:pipeline}
    The optimization pipeline behind the quantum solver. In the \emph{input stage}, a user-defined representation of the optimization problem is provided to the solver. Next, we construct the variational ansatz that is employed during the optimization process. In the \emph{compilation stage}, the variational circuit is compiled to produce an optimized parametric circuit. The compiler may leverage an enhanced gate set of pre-calibrated pulse-level instructions of efficient (fast) 2-qubit gates. Finally, the optimization parameters are initialized and then passed to the closed-loop optimizer. Depending on the number of circuit parameters, a parameter reduction procedure may take place. In the \emph{hybrid execution stage}, the parameterized circuits are submitted to the quantum hardware and the variational circuit parameters are optimized via a hybrid quantum-classical optimization loop. In the \emph{output stage}, after final post-processing, the solution to the optimization problem is returned to the user.}
\end{figure*}

We introduce an integrated hybrid quantum-classical optimization pipeline for binary combinatorial optimization problems on gate-model quantum computers that returns optimal or near-optimal solutions up to whole-chip 156-qubit implementations. We provide an overview of the quantum and classical components of the algorithm, including the use of a novel variational ansatz, a dual variational parameter update strategy using measurement-based feedback, and efficient classical post-processing to address bitflip errors. We benchmark this solver on IBM quantum computers using Q-CTRL parametric compilation and error suppression for several unconstrained binary optimization problems and compare against published results using alternate hardware systems. First, we consider Max-Cut instances \cite{Commander2009} for random regular graphs up to degree-7. Our implementation returned the correct Max-Cut value over all instances tested up to 156 nodes which were mapped directly to 156 qubits. We provide a quantitative comparison to previously published trapped-ion results using identical problem instances \cite{shaydulin2023qaoa}. Next, we applied the algorithm to higher-order unconstrained binary optimization (HUBO) problems, searching for the ground state energy of device-spanning random-bond spin-glass models with linear, quadratic, and cubic terms following Ref.~\cite{pelofske2023quantum}. For a 127-qubit problem, we demonstrated the ability to find the correct energy minimum in four of six instances tested, including an instance where the annealer did not find the correct ground-state energy; for the two additional instances tested, solution approximation ratios exceeded 99.5\%. For a 156-qubit problem, our algorithm found the global minimum on all instances considered. For both problem types, the algorithm outperformed a heuristic local solver that applies a greedy optimization to a random distribution of candidate solutions. Nonetheless, these problems are modest enough in scale such that they may be solved exactly using classical methods; we used CPLEX to provide the ground-truth. Overall, these results represent some of the largest quantum optimizations returning optimal solutions performed to date with on-device angle optimization for two problem classes that challenge classical solvers.

\section{Integrated Pipeline for Quantum Optimization}

We introduce a hybrid quantum-classical algorithm for unconstrained binary optimization problems over length-$n$ bitstrings of the form
\begin{equation}
    \label{eq:optimization_problem}
    \min_{\bm{x} \in \{0,1\}^n} C(\bm{x}),
\end{equation}
where the objective function $C(\bm{x})$ may be represented as a polynomial over the binary variables. The key elements of this algorithm, combined with Q-CTRL's compilation and error suppression tools, are visually represented in Fig.~\ref{fig:pipeline}. The algorithm builds upon on a commonly used quantum optimization algorithm, the Quantum Approximate Optimization Algorithm (QAOA), which is a hybrid classical-quantum algorithm inspired by the adiabatic theorem to approximate solutions for combinatorial optimization problems \cite{farhi2014quantum}.

After defining the optimization problem as in in Eq.~\ref{eq:optimization_problem}, the variational circuit is constructed as a standard QAOA implementation with cost and mixer Hamiltonians parameterized by $\gamma_i$ and $\beta_i$ where $i=1,...,p$ and $p$ represents the number of QAOA layers used. However, a change is made to the initial state, using additional parameterization that deviates from the initial layer of Hadamard gates in standard QAOA and replacing them with $R_y(\theta_j)$ gates, where $j=1,...,n$ represents the qubit index. All $\theta_j$ angles are initialized to $\pi/2$, recovering the standard QAOA ansatz at the beginning of the optimization procedure. 

The parameterized circuits are iteratively submitted to the quantum hardware as part of the hybrid quantum-classical optimization loop in which only the $\gamma_p$ and $\beta_p$ QAOA parameters are optimized with a closed-loop classical optimizer using the results from quantum circuit executions in order to iteratively minimize the cost function. The initial state parameters are updated at fixed points during the optimization process as detailed below to bias the initial state toward the best solution found over prior iterations. Execution of each circuit is performed using Q-CTRL's parametric compilation and error suppression pipeline detailed in Ref.~\cite{mundada2023experimental} in order to reduce gate-level and circuit-level errors. Finally, we post-process the measured distributions with a simple $O(n)$ greedy optimization pass. We review the key elements of the optimization algorithm pipeline in detail below.

\textit{Modified variational ansatz}: We aim to overcome the challenge of limited accessible circuit depth in commercial quantum processors, as this curtails the achievable QAOA repetition level, denoted $p$, well before the value $p>11$ anticipated to be required (at a minimum) to approach definitive quantum advantage~\cite{lykov2023sampling, boulebnane2022} in a conventional implementation. We follow previous literature to address both scalability and generalizability~\cite{bleckos2024review} through use of a modified QAOA ansatz and a dual variational parameter update strategy.

\begin{figure}[b!]
    \centering    \includegraphics[width=0.48\textwidth]{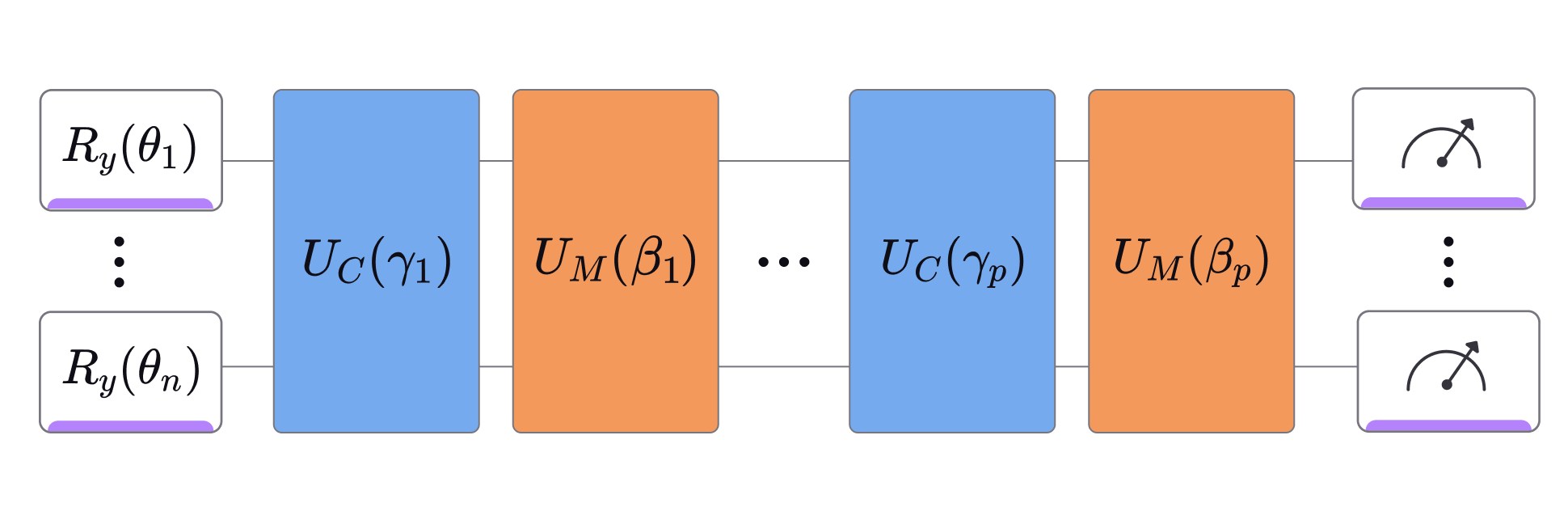}
    \caption{
    \label{fig:ansatz}
      The ansatz we employ for solving unconstrained optimization problems. The initial state is parameterized angles $\theta_j$, for $j=1,..,n$, followed by the standard QAOA repetition pattern of a cost-unitary and a mixing-unitary, defined by the $\gamma_i,\, \beta_i$ parameters for $i=1,...,p$. This ansatz has a total of $n+2p$ parameters where $n$ is the number of qubits and $p$ the number of QAOA layers.
      }
\end{figure}

The standard QAOA ansatz starts with an initial $n$-qubit equal superposition state established by application of a Hadamard gate to each qubit, followed by alternating layers of unitary transformations corresponding to the cost and mixer Hamiltonians. The cost Hamiltonian, $H_C$, encodes the optimization problem to be solved and the corresponding unitary transformation, $U_{C}$, is parameterized by variational parameters $\gamma_i$, where $i=1,\dots,p$. The mixer Hamiltonian, $H_M$, facilitates transitions between states in order to explore the solution space, and the mixing unitary $U_{M}$ for each layer is tuned with the variational parameters $\beta_i$.

Our modified QAOA ansatz, shown in Fig.~\ref{fig:ansatz}, substitutes the typical equal-superposition starting state with a new input state generated using arbitrary rotation operators $R_y(\theta_j)$, where $j=1,\dots, n$ and $\theta_j$ varies individually for each qubit. This parameterization is used to encourage convergence for standard QAOA by gradually biasing the optimization search space toward low-cost solutions found during the optimization process. All optimizations start with all $\theta_j$ set to $\pi/2$, producing the uniform superposition state, which is effectively identical to a standard QAOA ansatz. Our approach varies in that at specific intervals during the optimization process, we use this parameterization to bias the initial state toward the best solution observed during previous iterations. This adaptive approach builds on related work, but constitutes a departure from warm-start QAOA methods~\cite{egger2021warm, tate2023warm, truger2023warm, tate2023bridging} in which the initial state remains unchanged throughout the circuit optimization. Instead, our ansatz starts from the zero-information point in the first cycle of the classical loop, and the initial state parameters are updated two to three times during the optimization process, while $\{\bm{\gamma}, \bm{\beta}\}$ are optimized in every iteration as in a typical application of QAOA. The additional parameterization in this ansatz and parameter update strategy enables the solver to converge on correct solutions without the need for large $p$ by reducing the solution space through biasing at select intervals during the optimization. Such an approach mitigates practical concerns around the impact of incoherent errors and limited circuit depths encountered on today's machines. More details are provided in Appendix~\ref{app:initialstate}.

\textit{Efficient parametric compilation}: Our solver employs an efficient form of parametric compilation in order to reduce execution overheads---a practical rather than fundamental consideration that can have substantial impact on the achieved solution quality. During operation of the solver, the parameters of the executed quantum circuits are updated in real-time based on feedback from the classical optimizer during execution of the hybrid loop. This procedure repeats itself anywhere from tens of times for small problems to thousands of times for more complex problems.  Compilation overhead causes effective dead-time in which the hardware system is not being sampled, introducing sensitivity to drifts on the timescale of the optimization loop's execution. Accordingly, inefficient compilation introduces additional sensitivity to errors, the cumulative effect of which can impair convergence.

Our parametric compilation strategy incurs the initial compilation overhead only once, followed by real-time updates of the circuit parameters during the execution of the hybrid loop to minimize dead-time. Successful implementation requires careful consideration of the fact that symmetries and circuit identities that apply to specific parameters do not hold in general \cite{Future_compilation}. Additionally, while the available native gate set is universal, commonly used ansatze such as that in Fig.~\ref{fig:ansatz} may be compiled more efficiently by enriching the available gate set. In particular, our compiler can leverage pulse-level instructions of pre-calibrated gates to realize faster and higher quality arbitrary two-qubit rotations \cite{gate_eff_future}.

\textit{Automated error suppression in hardware execution}: The solver uses an automated error-suppression pipeline~\cite{mundada2023experimental} during hardware execution of each parameterized circuit. This strategy combines dynamical decoupling embedding for simultaneous cross-talk and dephasing suppression~\cite{coote2024dd} and automated AI-driven gate-waveform replacement~\cite{mundada2023experimental}. This pipeline has been shown to deliver $>1000\times$ performance enhancement in algorithmic benchmarks on IBM hardware~\cite{mundada2023experimental, qctrl_benchmarks}. In this work, we focus on the use of the modified variational ansatz and our results for two key classes of unconstrained optimization problems. Further details of the error suppression pipeline used in this work can be found in the cited references above.

\textit{Automated classical optimization loop}: We implement the classical component of the optimization loop using a covariance matrix adaptation evolution strategy (CMA-ES); it is a fast, stochastic, and derivative-free method used for non-convex numerical optimization problems \cite{Auger2005}. We select Conditional Value-at-Risk (CVaR) as the objective function, as it has been shown to lead to faster convergence for combinatorial optimization problems \cite{barkoutsos2020improving}. CVaR provides a weighted average of the tail end of a distribution; it is evaluated for the results of each circuit execution and CMA-ES uses these values to define the parameters for the variational circuits of the subsequent optimization step. The CVaR $\alpha$ parameter is the percentage of the tail used. For all data presented in this work, $\alpha=0.35$ was used during the optimizations.

To maximize the efficiency of quantum-resource utilization, we use multiple parameter update strategies in the optimization procedure. The number of variational parameters (the 2$p$ parameters $\bm{\gamma}$ and $\bm{\beta}$) is reduced by employing the Fourier parameterization introduced in Ref.~\cite{zhou2020qaoa}, although for the problems presented in this work $p=1$ was sufficient to find the known correct solutions (in which case, the Fourier parameterization is equivalent to the standard parameterization). Additionally, the $\bm{\theta}$ angles are initialized to $\pi/2$ (recovering the standard QAOA ansatz) and are updated a maximum of three times during the optimization run according to the information collected during the previous optimization steps; this procedure takes advantage of the bitstring distributions observed in the prior optimization steps to select the lowest-cost bitstring observed and bias the initial state toward it, with progressively stronger biasing strength. Every time the $\bm{\theta}$ angles are updated, we include a $(\bm{\gamma}, \bm{\beta})=(\bm{0},\bm{0})$ circuit, equivalent to a $p=0$ circuit, as a baseline comparison (see Appendix~\ref{app:initialstate} for details).
The $\bm{\gamma}$ and $\bm{\beta}$ parameters are initialized randomly; this method does not require problem-specific parameter initialization in order to converge toward high-quality solutions~\cite{akshay2021parameter}.

\textit{Classical post-processing}: Uncorrelated bitflip errors in the measured output distribution significantly impact solution quality. These errors can come from a variety of sources, commonly dominated by device $T_{1}$ noise and measurement errors. For large-scale problems beyond 100 qubits, eliminating these errors completely during circuit execution is unrealistic with contemporary devices. However, these errors may be addressed by a purely classical post-processing step.

We implement an $\mathcal{O}(n)$ na\"\i ve greedy optimization on the solution bitstrings returned from measuring the final, optimized variational circuit. This may be done efficiently given that the cost function $C$ is local---evaluating the cost difference $C(\bm{x}') - C(\bm{x})$ for $\bm{x}, \bm{x}'$ differing by a single bitflip involves a constant number of steps. With no additional measurements or circuit modifications required, this procedure randomly traverses the measured solution bitstring and greedily flips bits if they improve the bitstring cost. This approach continues until no improvement is realized from successive bitflips, and we cap the number of full traversals of the bitstring to five passes. This implementation is not problem-dependent and can be applied to any unconstrained binary combinatorial optimization problem. Importantly, it requires no additional execution overhead in increased shots on hardware. The use of greedy classical optimization in conjunction with QAOA has been explored in Refs. \cite{dam2021quantum, caha2022twisted, sack2023recursive, dupont2024quantum, wurtz2024solving}.

This post-processing pass is not specifically a bitflip error-correcting pass but a computationally cheap optimization to provide some level of robustness to uncorrelated bitflip errors that are unavoidable on current noisy devices, especially in large-scale problems.

\section{Hardware implementation}

We test the performance of the solver via execution on the 127-qubit IBM quantum computers \textit{ibm\_brisbane} and \textit{ibm\_sherbrooke}, as well as the 156-qubit \textit{ibm\_kingston} and \textit{ibm\_boston} devices. The data on 127-qubit IBM Eagle devices was collected in 2024, and the data on 156-qubit IBM Heron devices was collected in September 2025 (Max-Cut) and January 2026 (spin-glass instances). In all cases, the circuit executions within the full optimization loop are executed on the quantum device, without any use of classical simulation to optimize parameters. The method implementation is not tuned or modified for the specific problems we implement aside from the hardware-efficient spin-glass problem embedding, and the results we show faithfully represent the quality of expected solutions for any optimization problem in the same class as the ones we optimized.

We evaluate performance by determining a series of metrics which highlight the quality of the output solutions relative to the global minimum (or minima, in the case of degenerate solutions), and the likelihood of achieving the correct solution. In our tests, the global minimum is determined using the classical solver CPLEX \cite{cplex}. First, we evaluate the approximation ratio; for a solution defined over binary variables as $\bm{x}\in \{0,1\}^n$, the approximation ratio is
\begin{equation}
    \label{eq:approximation_ratio}
    \text{AR}(\bm{x}) = \frac{C(\bm{x}) - C_{\text{max}}}{C_{\text{min}} - C_{\text{max}}},
\end{equation}
where $C(\bm{x})$ is the objective function, $C_{\text{min}}$ and $C_{\text{max}}$ are the minimum and maximum objective value of the problem (both may be obtained via CPLEX, the maximum corresponds to minimum objective for the negated problem $-C(\bm{x})$). An AR$\;=100\%$ corresponds to the algorithm finding a solution matching the minimum.

In addition, we provide standard metrics allowing quantitative comparison of solution quality by including the likelihood that the top solution is returned. This metric is particularly relevant for cases where the achieved AR$\;=100\%$, i.e., the top solution is the correct solution, as it provides additional information on the stability of the optimization process and the relative correctness of each sampling of the optimized circuit. Any circuit, when sampled, will return a distribution of outputs that ideally skews towards the ground truth; a high-quality solver will deliver a large overlap of outputs with the ground truth while a weaker solver may find the ground-truth solution only rarely if at all.

When available, we compare the quality of solutions returned from the solver against published results derived from a commercial D-Wave annealer and a Quantinuum trapped-ion gate-model quantum computer. In such instances, the problems we execute are defined identically to those executed in the relevant references. In addition, to provide insight into the relative challenge of the test problems we explore, we also include a classical local solver in our benchmarks. A summary of all results achieved with comparisons against previously published literature appears in Table~\ref{table:results}. Additional information including estimates of execution time and cost is included for completeness in Appendix~\ref{app:resources}.

\begin{figure*}[t!]
    \centering
    \includegraphics[width=1.0\textwidth]{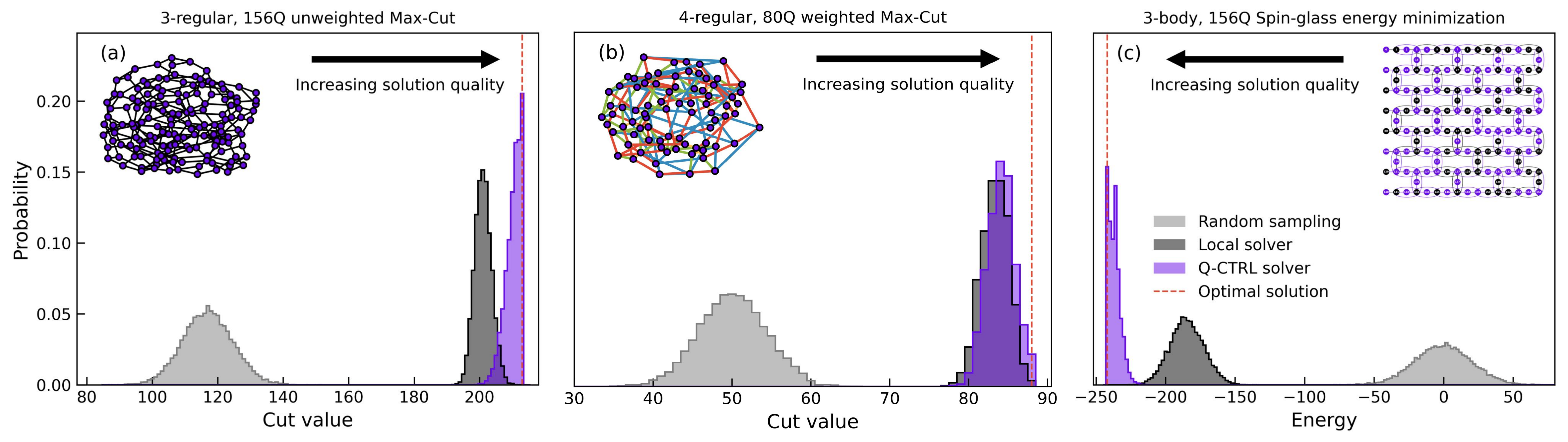}
    \caption{
    \label{fig:maxcut_and_lanl_combined_plot}
    Performance of the quantum solver on three combinatorial optimization problems.
    (a) A Max-Cut instance for a 3-regular unweighted random graph with 156 nodes. The plot shows the distribution of cut values for 15k configurations sampled from the optimal circuit found by the quantum solver, executed on \textit{ibm\_kingston}. The true maximum cut value of 213 is successfully attained by the solver. The random search and classical local solver distributions show the result of 15k configurations sampled uniformly at random and 15k local minima.
    (b) A Max-Cut instance for a 4-regular weighted random graph with 80 nodes. The weight of each edge was randomly chosen from four possible values $[1/4, 1/2, 3/4, 1]$, represented by different edge colors. Shown, cut values for 10k configurations sampled from the optimal circuit found by the quantum solver, executed on \textit{ibm\_brisbane}. The true maximum cut value of 88 is successfully attained by the solver.
    (c) Energy minimization of a 156-node high-order spin-glass model with parameters defined from \textit{ibm\_kingston} instance 0 following similarly defined problems in Ref.~\cite{pelofske2024repo}. The plot shows the energy distribution for 15k configurations sampled from the optimal circuit found by the quantum solver, executed on \textit{ibm\_kingston}, and for 15k configurations sampled uniformly at random (brute-force) and 15k local minima (local solver). The lowest and highest energies of this model instance (the energy band edges) are known to be $-242$ and $242$, respectively. As in (a,b), the true ground state was attained by the solver.
    }
\end{figure*}

\subsection{Max-Cut}
The Max-Cut problem is a classic problem in combinatorial optimization and graph theory. It involves partitioning the vertices of a graph into two distinct subsets such that the number of edges between the two subsets is maximized. There is a formal equivalence between Max-Cut and QUBO problems---any Max-Cut instance may be mapped to a QUBO instance and vice versa \cite{barahona1989experiments}. Max-Cut on $k$-regular graphs, meaning graphs where each node is connected to exactly $k$ other nodes, are NP-hard problems in the case where $k>2$ and have real-world significance, particularly in networking \cite{Commander2009, Jumaa_2020}. For arbitrary graphs, this problem is also challenging because it does not exploit native symmetries or connectivity of the underlying quantum hardware in use. Accordingly, Max-Cut on random 3-regular graphs is a standard benchmark problem used to test the efficacy of quantum algorithms against classical counterparts by being both a simple and nontrivial example of a QUBO problem.

Max-Cut can be framed as a minimization of an objective function of the form
\begin{equation}
    C(\bm{z}) = -\frac{1}{2} \sum_{i < j} A_{ij} \left(1 - z_i z_j \right) \,.
\end{equation}
Here, we have expressed the cost function in terms of the Ising spin variables $\bm{z} = 2 \, \bm{x} - 1 \in \{-1, 1\}^n$, and $A$ is the $n \times n$ adjacency matrix of the graph ($A_{ij} \neq 0$ if $(i,j) \in G$, otherwise $A_{ij} = 0$), with $n$ is the number of nodes. We distinguish between unweighted graphs where $A_{ij} = 1$ if $(i,j) \in G$, and weighted graphs where $A_{ij} = w_{ij} \in \mathbb{R}$ if $(i,j) \in G$. For all problems we present, the number of nodes is equivalent to the number of qubits used on the device. Candidate partitions of the graph are then represented by length-$n$ bitstrings (entries are zero or one depending on which subgraph each node is assigned to), and each graph node is assigned to a single qubit. Given a partition, the cut value is calculated straightforwardly as the number of edges between the two groups.

In Fig.~\ref{fig:maxcut_and_lanl_combined_plot}(a-b), we demonstrate the performance of our solver on two example Max-Cut problems: An unweighted 156-node 3-regular graph and a weighted 80-node 4-regular graph (see Table~\ref{table:results} for summary results on these and additional instances). In these figures, we plot the histogram of cut values calculated from the bitstring solutions returned by the solver, output as measurements of the optimal circuit. A good solution approaches and ideally overlaps the actual maximum value derived from CPLEX (denoted by a dashed vertical red line) towards the right side of the plot.

\begin{figure*}[t!]
    \centering
    \includegraphics[width=1.0\textwidth]{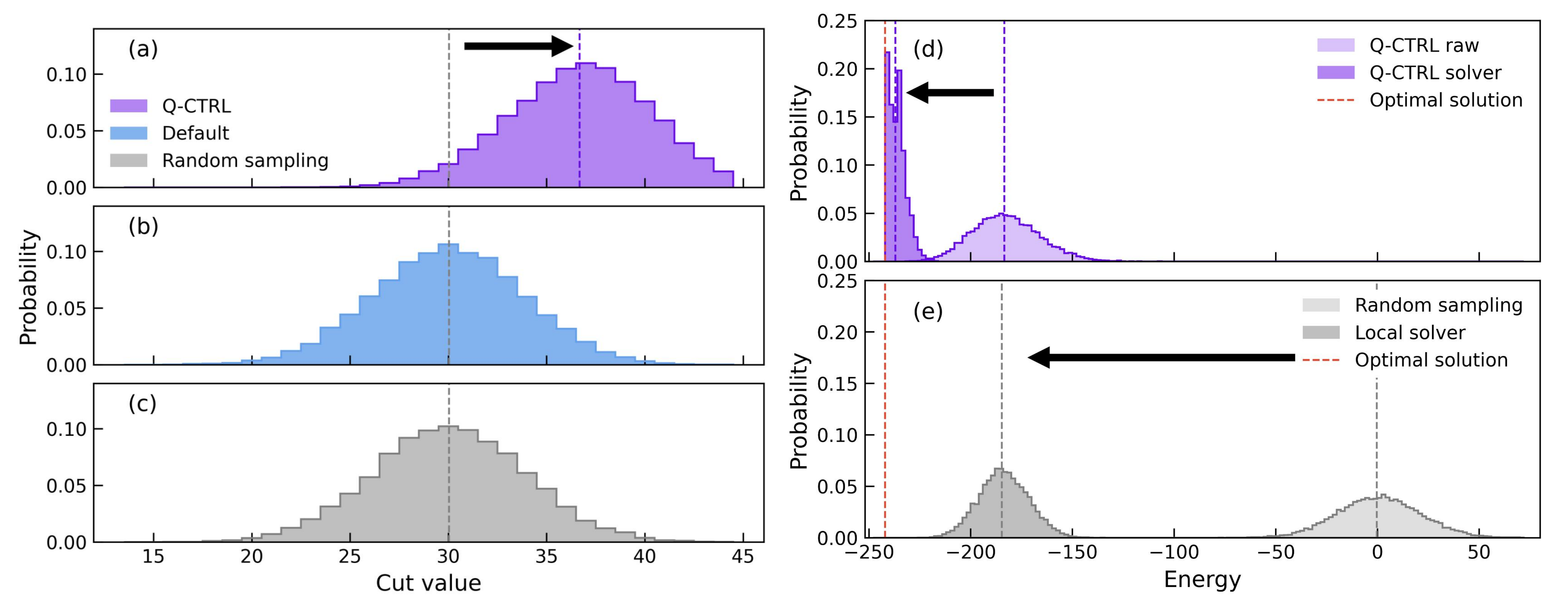}
    \caption{
    \label{fig:error_suppression_and_postprocessing}
    (a-c) The role of error suppression. The distribution of cut values obtained by sampling 16,384 configurations from a $p=1$ QAOA circuit for a Max-Cut instance for a randomly generated 3-regular graph with $40$ nodes. %
    The circuit was executed on the \textit{ibm\_sherbrooke} device utilizing two performance management schemes, (a) using Q-CTRL error suppression pipeline and (b) Qiskit version 0.44, with optimization level 3 and resilience level 1. In all executions, the equal weights superposition state served as an initial state, optimal QAOA parameters were used and the data shown does not include post processing. (c) A reference distribution corresponding to uniformly sampling bitstrings at random.
    The purple and gray dashed lines indicate the mean of the Q-CTRL and the reference distribution.
    (d-e) The role of post processing. (d) The distributions of cost values obtained from the optimal circuit returned by our pipeline, before and after the application of post-processing, for cubic spin-glass instance 0 run on \textit{ibm\_kingston} also shown in Fig.~\ref{fig:maxcut_and_lanl_combined_plot}c.
    (e) For reference, we show the distributions of cut values obtained by 15k random bitstrings and by applying post-processing on each of these 15k configurations, denoted as "local solver."
    }
\end{figure*}

Our performance baseline is set by sampling bitstrings uniformly at random (light gray histogram), akin to a random brute-force scan through the solution set. In all cases, the number of samples is taken to be the same as the samples of the optimal QAOA circuit. We choose this baseline as an indication of the difficulty of the problem and also because, as shown later, a na\"\i ve QAOA implementation using standard Qiskit compilers returns results indistinguishable from random selection (see Fig.~\ref{fig:error_suppression_and_postprocessing} for detail).

Executing the entire quantum-solver workflow depicted in Fig.~\ref{fig:pipeline} results in a shift of the output histogram to large cut values, and produces solutions which attain the known maximum cut value. Up to 156 qubits, we find that not only does the quantum solver distribution overlap the maximum cut value, but it is asymmetrically skewed towards this value resulting in a high likelihood of achieving the correct answer. The quantum solver also provides superior solutions in these cases to those found using a classical local solver shown in the dark gray histogram (see Appendix~\ref{app:classical} for implementation details).

In Table~\ref{table:results}, we show additional results across different Max-Cut problems with higher-density graphs, up to 7-regular graphs. In all executions attempted, and across all specific problem instances within each class, the quantum solver achieves AR$\;=100\%$. In these cases, the solver also uniformly outperforms the classical local solver in solution likelihood, and matches the CPLEX ground-truth solutions. Additional data across the problem instances explored appear in Appendix~\ref{app:data}.

These results demonstrate superior performance to the previously published results on a trapped-ion quantum processor with 32 qubits in Ref.~\cite{shaydulin2023qaoa}. Because we do not have access to the same trapped-ion device, we cannot comment on relative base-hardware performance had the Q-CTRL pipeline been implemented; rather we only make comparisons to the results published to date. Table~\ref{table:results} provides a direct comparison of executions of the quantum solver on IBM hardware using the same problem instances presented in \cite{shaydulin2023qaoa} for smaller graphs up to 32 nodes. Our solver achieves AR$\;=100\%$ across all instances for $p=1$. In execution of higher-depth circuits with $p=10$ and $p=11$, the trapped-ion device returned the maximum cut, but with up to $9\times$ lower likelihood than our solver; across these problem instances our solver finds the maximum cut with likelihood of $\sim$82--94\%, while on Quantinuum the range was $\sim$10--18\%. Perhaps more importantly, the trapped-ion demonstrations were limited to hardware execution of a single circuit using pre-optimized $\bm{\gamma}$ and $\bm{\beta}$ parameters attained via classical numeric simulation. Our solver, by comparison, exclusively runs full hardware execution of the QAOA optimization loop. However, we note that there exists another trapped-ion demonstration of a full optimization of $p=2$ QAOA for a 32-node Max-Cut problem that returns the optimal cut \cite{moses2023racetrack}.

\begin{table*}[t!]
\begin{center}
\begin{tabular*}{\textwidth}{@{\extracolsep{\fill}}cc|cccc|ccc|ccc} %
\toprule
\multicolumn{12}{l}{\textbf{Max-Cut}} \\
\midrule
\multicolumn{2}{l}{\textbf{Global}} & \multicolumn{4}{|l|}{\textbf{Q-CTRL (IBM Quantum)}} & \multicolumn{3}{l|}{\textbf{Local solver}} & \multicolumn{3}{l}{\textbf{Quantinuum (H2-1)}} \\
\multicolumn{2}{l}{} & \multicolumn{4}{|l|}{Full hybrid optimization} & \multicolumn{3}{l|}{} & \multicolumn{3}{l}{Classically optimized in simulation} \\
\midrule
Instance & Min/Max & Max & AR (\%) & Likelihood & Device & Max & AR (\%) & Likelihood & Max & AR (\%) & Likelihood \\
\midrule
(28, 3, 102, u)   & 0 / 40    & \textbf{40}    & \textbf{100} & 9.44E-1 & S & \textbf{40}    & \textbf{100} & 6.20E-1    & \multicolumn{1}{c}{\textbf{40}} & \multicolumn{1}{c}{\textbf{100}} & 1.78E-1 \\
(30, 3, 264, u)   & 0 / 43    & \textbf{43}    & \textbf{100} & 9.18E-1 & S & \textbf{43}    & \textbf{100} & 4.68E-1    & \multicolumn{1}{c}{\textbf{43}} & \multicolumn{1}{c}{\textbf{100}} & 1.04E-1 \\
(32, 3, 7, u)     & 0 / 46    & \textbf{46}    & \textbf{100} & 8.25E-1 & S & \textbf{46}    & \textbf{100} & 4.06E-1          & \multicolumn{1}{c}{\textbf{46}} & \multicolumn{1}{c}{\textbf{100}} & 1.27E-1  \\ \cmidrule{10-12}
(80, 3, 68, u)    & 0 / 106   & \textbf{106}           & \textbf{100}  & 1.38E-1 & Br & \textbf{106}   & \textbf{100} & 1.00E-4  &                                 &                                  &                                  \\
(100, 3, 12, u)   & 0 / 135   & \textbf{135}   & \textbf{100} & 1.21E-1 & Br & \textbf{135}   & \textbf{100} & 1.00E-4          &                                 &                                  &                                  \\
(120, 3, 8, u)    & 0 / 163   & \textbf{163}   & \textbf{100} & 8.59E-2 & Br & \textbf{163}   & \textbf{100} & 1.67E-4 &                                 &                                  &                                  \\
(156, 3, 0, u) & 0 / 213    & \textbf{213}    & \textbf{100} & 2.05E-1 & K & 211    & 94.4 & 1.33E-4           &                                 &                                  &                                  \\
(156, 5, 0, u) & 0 / 324    & \textbf{324}    & \textbf{100} & 2.13E-3 & K & 322    & 95.4 & 6.67E-5           &                                 \multicolumn{3}{c}{Not applicable}                                      \\
(50, 6, 28, w) & 0 / 73.2  & \textbf{73.2}  & \textbf{100} & 8.10E-2 & S & \textbf{73.2}  & \textbf{100} & 3.67E-2          &                            &  &       \\
(50, 7, 27, w) & 0 / 86.25 & \textbf{86.25} & \textbf{100} & 7.71E-2 & S & \textbf{86.25} & \textbf{100} & 3.67E-2          &                                 &                                  &                                  \\
(60, 5, 36, w) & 0 / 76.25 & \textbf{76.25} & \textbf{100} & 1.01E-1 & S & \textbf{76.25} & \textbf{100} & 1.57E-2          &                                 &  &                           \\
(70, 4, 75, w) & 0 / 83.75 & \textbf{83.75} & \textbf{100} & 6.16E-2 & S & \textbf{83.75} & \textbf{100} & 2.20E-3          &                                 &                                  &                                  \\
(80, 4, 46, w) & 0 / 88.0    & \textbf{88.0}    & \textbf{100} & 2.08E-2 & S & \textbf{88.0}    & \textbf{100} & 2.90E-3           &                                 &                                  &                                  \\

\midrule
\multicolumn{12}{l}{\textbf{Spin Glass}} \\
\midrule
\multicolumn{2}{l}{\textbf{Global}} & \multicolumn{4}{|l|}{\textbf{Q-CTRL (IBM Quantum)}} & \multicolumn{3}{l|}{\textbf{Local solver}} & \multicolumn{3}{l}{\textbf{D-Wave (Pegasus)}} \\
\multicolumn{2}{l}{} & \multicolumn{4}{|l|}{Full hybrid optimization} & \multicolumn{3}{l|}{} & \multicolumn{3}{l}{Grid search} \\
\midrule
Instance & Min/Max & Min & AR (\%) & Likelihood & Device & Min & AR (\%) & Likelihood & Min & AR (\%) & Likelihood \\
\midrule
0  & -200 / 192 & \textbf{-200} & \textbf{100}         & 1.34E-4 & S & -188 & 96.9 &6.67E-5       & \textbf{-200} & \textbf{100} & 7.53E-7 \\
3  & -198 / 184 & \textbf{-198} & \textbf{100}         & 1.18E-1  & S &-190 & 97.9 & 6.67E-5 & \textbf{-198} & \textbf{100}&7.53E-7 \\
5  & -198 / 192 & \textbf{-198} & \textbf{100}         & 2.03E-1  & S & -188 & 97.4 & 2.67E-4 & -194 & 99.5  &3.77E-6 \\ \cmidrule{10-12}
10 & -202 / 190 & \textbf{-202} & \textbf{100}         & 1.92E-2  & Br &-194 & 98.0 & 6.67E-5   &                                   &                                  &                              \\
11 & -180 / 196 & -178          & 99.5          & 1.69E-1 & Br & -170 & 97.3 & 6.67E-5  &                                    & &                               \\
69 & -190 / 212 & -188 & 99.5 & 1.75E-1 & Br & -180 & 97.5 & 6.67E-5   &                                   &                                  &                              \\ 
0 & -242 / 242 & \textbf{-242} & \textbf{100} & 2.17E-1 & K & -226 & 88.2 & 1.33E-4   &                                  \multicolumn{3}{c}{No published data}                            \\
0 & -242 / 242 &  \textbf{-242} & \textbf{100} & 1.33E-4 & Bo & -226 & 88.2 & 1.33E-4   &                                   &                                  &                              \\
1 & -240 / 238 & \textbf{-240} & \textbf{100} & 4.64E-1 & Bo & -228 & 88.2 & 6.67E-5   &                                   &                                  &                              \\
2 & -250 / 246 & \textbf{-250} & \textbf{100} & 1.67E-3 & Bo & -230 & 87.0 & 6.67E-5   &                                   &                                  &                              \\
\bottomrule
\end{tabular*}
\end{center}
\caption{
\label{table:results}
Comparison between different solvers on a variety of (top) Max-Cut problems and (bottom) spin-glass energy minimization problems. Max-Cut problems are specified by four parameters $(n,k,s,u/w)$, where $n$ is the number of graph nodes, $k$, the degree of the regular graph, $s$, the seed of the random graph and u/w indicating if the graph is unweighted (u) or weighted (w). The seeds for the first three problems define the problems accessed from Ref.~\cite{shaydulin2023qaoa_data}. The other Max-Cut problems are defined by the seeds used to generate them with NetworkX's random regular graph generator \cite{networkx}. The spin-glass problems are specified by instance numbers. For the first 6 entries, these correspond to the same instance numbers used in \cite{pelofske2024repo} for the problems generated for the \textit{ibm\_washington} device (note that in this work, these problems were executed on other devices). For the last 4 entries, the instance number corresponds to the random seed used to generate the problem instance. We use the same circuit construction approach as \cite{pelofske2023scaling} for the spin-glass problems to facilitate a fair comparison. Exact global minimum and maximum values were evaluated using CPLEX. For each solver, we list the best objective found (top solution), its approximation ratio (AR), and the observed top solution likelihood (estimated as the ratio of the number of times it was sampled to the total number of samples). The integrated solver data indicates the top solution found out of $N_{\text{shots}}$ used for sampling the optimal circuit. For all spin-glass problems $N_{\text{shots}} = 15$k, where for the Max-Cut problems the number of shots varies between 6--15k depending on the number of nodes (see Appendix~\ref{app:data} for more experimental details). The IBM device used for each problem is indicated as Br for \textit{ibm\_brisbane} (127Q), S for \textit{ibm\_sherbrooke} (127Q), K for \textit{ibm\_kingston} (156Q), and Bo for \textit{ibm\_boston} (156Q). The local solver data indicates the top solution found out of $N_{\text{shots}}$ local minima found by a greedy algorithm applied to a random distribution of candidate solutions (see Appendix~\ref{app:classical} for implementation details). The QA data, taken from \cite{pelofske2024short}, indicates the top solution found out of $1,328,000$ samples collected from two Pegasus-class D-Wave devices. The Quantinuum data, taken from \cite{shaydulin2023qaoa_data}, is based on 1024 samples of a pre-optimized QAOA circuit, executed on the H2-1 device, where for each problem the best performing $p$, out of $p\leq11$, is shown.
}
\end{table*}

\subsection{HUBO Ising spin glass}
Next, we consider an optimization problem designed to find the ground-state energy of a random-bond Ising model possessing cubic terms. This is an example of a Higher-order Unconstrained Binary Optimization (HUBO), known to be more challenging to solve than QUBO problems \cite{Yarkoni_2022, deng2023,verchere2023, Glos2022}.

In our benchmarking, we follow the same model as introduced in \cite{pelofske2023scaling}, adapted to two different IBM device architectures. We consider a spin Hamiltonian with 1-, 2-, and 3-body interaction terms tailored to the heavy-hex connectivity of (1) the Eagle-class devices with 127-qubits, and (2) the Heron-class devices with 156-qubits. Denoting the heavy-hex connectivity graph as ${G = (V, E)}$, the cost function for this problem is
\begin{equation}
\label{eq:spin_glass}
    C(\bm{z}) = \sum_{i \in |V|} a_i z_i + \sum_{(i,j) \in E} b_{i,j} z_i z_j + \sum_{(i,j,k) \in W} c_{i,j,k} z_i z_j z_k \,.
\end{equation}
The coefficients $a$, $b$, and $c$ represent the 1-, 2-, and 3-body coupling strengths, respectively, with $W$ denoting a specific set of 3-tuples corresponding to 3-body interaction terms optimized for efficient implementation on IBM hardware. For Eagle-class devices, the resulting objective function comprises 127 linear terms, 142 quadratic terms, and 69 cubic terms, while Heron-class devices support expanded problem instances with 156 linear terms, 176 quadratic terms, and 244 cubic terms. The values of the coefficients for each term are randomly chosen to be $\pm 1$. We refer to different choices of these coefficients as model instances, and use identical instances and notation as in \cite{pelofske2023scaling} for the problems defined on the \textit{ibm\_washington} device topology.

In Fig.~\ref{fig:maxcut_and_lanl_combined_plot}c we show the performance of our solver for ``instance 0'' defined on \textit{ibm\_kingston} (see Appendix~\ref{app:data} for the remaining instances). Here, the distribution produced by the quantum solver is narrowly concentrated around the edge of the energy band (in this case solutions at the left edge of the plot are best), producing high quality ground-state estimates that match the known ground state energy. As above, this solution outperforms all candidate solutions returned by random sampling. As for Max-Cut, we also benchmark against a local solver, showing that our solver consistently outperforms the local solver. We provide further detail in Table \ref{table:results} where we highlight the performance of the solver over multiple problem instances reproduced from~\cite{pelofske2024short}. The worst-case approximation ratio returned for the selected problem instances tested was AR$\;=99.5\%$.

Through the introduction of additional variables and interaction terms in the energy, this HUBO problem may be converted to a QUBO problem, allowing for execution on quantum annealers \cite{Yarkoni_2022}. A $127$-node HUBO can be written as a $265$-node QUBO with $625$ quadratic terms and $69$ penalty constants. The nonlocal nature of the additional quadratic terms and the additional energy scales introduced by the multiple penalty terms render the extended QUBO problem challenging for quantum annealers \cite{VALIANTE2021}. For that reason, in order to benchmark this problem using an annealer, the authors in \cite{pelofske2024short} developed a dedicated order-reduction scheme that was tailored for the specific problem and for the specific QA device used in that work.

In Ref.~\cite{pelofske2024short}, quantum annealing was benchmarked for ten different problem 127-qubit instances. The instance above (number 5) was the only one where the ground state was not sampled at all. Yet, in other instances where AR$\;=100\%$ for both the quantum solver and the annealer, success on the annealer was marginal. For example, in two instances the correct ground state was measured only once out of 1,328,000 samples, corresponding to success likelihood of $\sim7.5\times 10^{-7}$ (see instances 0 and 3 in Table~\ref{table:results}). By contrast for these same instances the quantum solver sampled the solution with likelihood of $\sim 1.3\times 10^{-4}$ and $\sim 0.118$. We note that this comparison involves different optimization strategies: the quantum annealing study employed grid search over annealing parameters, while our approach optimizes variational circuit parameters, making a direct methodological equivalence challenging to establish. %

\section{Benchmarking Performance Analysis}
The results presented above indicate that it is possible to achieve useful and correct results from quantum optimization executed at full device scale on gate-model superconducting quantum computers. The ability to consistently solve complex problems at large scales, as we demonstrated, results from the combination of performance enhancements delivered via various parts of the full solver pipeline. The role and contribution of each subroutine strongly depends on the optimization problem; below we include experimental comparisons in order to provide qualitative insights regarding the role of each component. %

Information extracted during the quantum circuit execution is the core difference between the quantum solver and classical counterparts. Therefore, the ability to faithfully execute circuits on the quantum hardware is directly translated to the value the quantum solver provides. We now discuss the essential role played by error suppression in ensuring high-quality outputs from the quantum solver. We build on previous demonstrations that the QAOA landscape could show up to $200\times$ greater structural similarity to the ideal when comparing default circuit implementation to that realized in our integrated error-suppressed quantum solver~\cite{mundada2023experimental}.

Fig.~\ref{fig:error_suppression_and_postprocessing}(a-c) shows the distribution of sampled circuit outputs for an example 40-qubit Max-Cut problem in which we isolate the hardware execution by removal of key parts of the pipeline including the modified QAOA ansatz and output post-processing. This data was collected in early 2024; here we compare our integrated solver's hardware execution to the default execution achieved using Qiskit (version 0.44, with optimization level 3 and resilience level 1) and to random sampling. These data indicate that default circuit execution without additional error-suppression technology is indistinguishable from random sampling, and thus the quantum nature of the optimization plays no useful role. By contrast, the use of error-suppression leads to a separation of the distributions towards larger cut values, indicating more useful outputs from sampling the quantum circuit. %

Next, we compare the quantum solver with and without post-processing to the classical local solver in order to disambiguate benefits coming from the quantum optimization and classical \emph{post-processing}. In Fig.~\ref{fig:error_suppression_and_postprocessing}(d-e), we show the performance of the quantum and local solvers with and without post-processing for the spin-glass optimization problem presented in Fig.~\ref{fig:maxcut_and_lanl_combined_plot}c.

In both the quantum and classical cases, post-processing plays a significant role. The quantum solver exhibits a shift of its distribution to large cuts and achieves the correct Max-Cut solution with $\sim8.5\%$ likelihood (with post-processing AR$\;~\sim82\to100\%$). The local solver (which includes postprocessing) also finds the correct solution after post-processing and sees a shift towards higher AR than simple random search, but with $\sim500\times$ lower likelihood than the quantum solver. This is visible from the fact that only the right-hand tail of the local solver's distribution overlaps the true maximum cut value, whereas the quantum solver's distribution is skewed towards the correct answer. Refs.~\cite{mathieu2008greedy, costello2011greedy} show that the AR achieved through a greedy classical optimization algorithm alone for Max-Cut is expected to be 0.5. Nonetheless, as a post-processing step, it enables our solver to achieve much faster convergence to high-quality solutions and reduces the sensitivity of the algorithm to uncorrelated random bitflip errors. We also use this greedy algorithm as part of our $\bm{\theta}$ update strategy.

Finally, we discuss the role of the modified variational ansatz and associated \emph{parametric compiler} employed in the quantum solver. The ability to individually vary the angles $\bm{\theta}$ allows the ansatz to begin in a nonuniform superposition that biases the QAOA circuit toward a low-cost solution, which we observe permits a reduction in the number of QAOA layers needed to converge on high-quality solutions. In our experiments, analysis of the convergence process indicates that the QAOA-variational-parameter optimization is necessary for convergence; i.e., we observe improvement from the quantum optimization beyond what is attainable in the $p=0$ limit (see Appendix~\ref{app:initialstate} for further discussion).

Our approach adopts a parameterized initial state to reduce circuit depth necessary to achieve convergence. Nonetheless, even in the shallow limit, the number of two-qubit gates in each QAOA block increases dramatically with the connectivity and scale of the problem.

Implementing the modified QAOA ansatz requires efficient compilation to handle large-scale circuits. The Q-CTRL parametric compiler effectively reduces both circuit gate counts and duration compared to other commercially available parametric compilers~\cite{Future_compilation, gate_eff_future}. In Appendix~\ref{app:compilation}, we compare the quality of compilation, measured in terms of the number of two-qubit entangling gates required for key sample problems. These include $p=2$ QAOA circuits of $n$-node Max-Cut problems over random $3$-regular graphs and a more generic parametric circuit structure \cite{Temme2019, ZZ_map}. Such structures are likely to appear for constrained optimization where the circuit ansatz may include non trivial mixing layers or more dense parameterization. The demonstrated compiler efficiency enables the greater performance of the optimization routine described above without incurring any substantive penalty in noise-susceptibility as might be expected by inclusion of the parameterized initial state to the ansatz.

\section{Conclusion}
In this manuscript we have introduced and benchmarked an end-to-end quantum solver for binary optimization problems that \emph{successfully} solves classically nontrivial problems at the scale of $156$ qubits -- the largest reported to date. The methods we introduce consistently found the known optimal solution for Max-Cut and HUBO spin-glass problems; worst-case problem instances delivered approximation ratios $\ge 99.5\%$. Moreover, when compared against the open literature for identical problem instances, these results demonstrate a factor of up to $9\times$ enhancement in the likelihood of finding the correct solution over the best published results using a trapped-ion gate-model machine and over 1,500$\times$ enhancement vs published results on an annealer. While these comparisons involve different optimization methodologies (variational parameter optimization versus grid search approaches), they represent the most direct performance assessments possible given available published data. This metric is material in that it translates directly into required runtime and hence user cost. The new quantum solver was able to handle a broad range of problem classes and topologies, and relied exclusively on hardware execution without any classical simulation or pre-computation involved.

The results presented in this work demonstrate that an appropriately constructed quantum solver executed on an IBM gate-model quantum computer can outperform published outputs from quantum annealers for nontrivial binary optimization problems. Such a statement warrants careful qualification. First, we do not comment on the potential for different relative performance among gate-model machines should our quantum solver be implemented on alternative hardware. Next, as we relied upon published problem instances and implementations for quantum annealers we do not exclude the possibility of crafting better solutions in detailed implementation. We note that the higher-order binary optimization implemented by the authors of~\cite{pelofske2024short} was highly tailored to the problem and carefully constructed to efficiently use the available hardware resources in mapping a HUBO problem to the QUBO problem class efficiently addressed by annealers. Nonetheless it contains no optimality proof and an alternate implementation could deliver superior annealer results. Further, our results do not preclude the existence of alternate problem statements in which annealers retain advantages over gate-model machines, or that individual hardware systems may undergo periodic upgrades~\cite{DWaveadvantage} to deliver improved performance relative to the reference results used here. Overall, despite these qualifications we believe these results now challenge the heretofore practically correct assertion that annealers would consistently outperform gate-model machines for quantum optimization problems.

These results also suggest the potential for significant advantages in future scaling of problem size and complexity leveraging the gate-model architecture. This is because the use of this architecture directly maintains the richness and flexibility to generalize to a broad class of problems beyond QUBO, as we show in the problems treated here. For the avoidance of any doubt we do not claim quantum advantage over or even practical equivalence with the highly efficient classical CPLEX solver. Nonetheless, the success of the quantum solver we have tested at full device scale provides a potential line of sight to solving much larger future problems which do challenge classical heuristics. Higher complexity problems lead to a higher gate count in the required circuits, but we believe increasing hardware scale and performance shows a potential path to quantum advantage in quantum optimization. We look forward to continued innovation in this space, including in novel alternate optimization algorithms that could be combined with our solver~\cite{wurtz2021classically,amaro2022filtering,cadavid2024,granet2024benchmarking}.

\subsection*{Comment on recent experimental demonstrations}
We comment on recent experimental demonstrations published during preparation of this manuscript. First, \cite{montanezbarrera2024} suggests signatures of quantum advantage on QAOA problems up to 109 qubits. The problems treated were defined on a simpler linear topology and hardware implementations were not successful finding the correct solutions above 30 nodes. Further, the achieved values AR$\;<80\%$ suggest that classical local solvers may be able to outperform these results (see Table~\ref{table:results_appendix}). Results in~\cite{cadavid2024} present an alternative method to QAOA, and deliver a roughly $30\%$ enhancement over a standard implementation in simulation. Hardware execution of a pure QUBO problem at 100-qubit scale achieved AR$\;\sim65\%$.

D-Wave provided a response to an earlier version of this manuscript in which they include an updated performance comparison to the Max-Cut and spin-glass results presented here \cite{dwave_comment}. We do not compare all of the results of our solver to QA as we did not have access to published data for all problem instances we treated, limiting our direct comparison to the category of high-order spin-glass problems. For such higher-order optimization problems, the D-Wave results published in Ref.~\cite{dwave_comment} confirm that the integrated solver outperforms annealers for two of the cubic spin-glass problem instances, even with their new and augmented implementation that also includes our post-processing pass.

D-Wave also offers a comparison in which the team claims annealers maintain an advantage using their definition of a time-to-solution (TTS) metric. In this manuscript, we present only a comparison of probability for the final circuit from the QAOA process to sample the correct solution (comparable to execution of a D-Wave Hamiltonian following the embedding procedure), and we provide information on both sample time and runtime for the process (Appendix~\ref{app:resources}). We note an apparent discrepancy between D-Wave's definition of sample time and the actual sample time on hardware which leads to a difference of $\sim100\times$ favoring annealers. The value of the sample time as defined therein is heavily dependent on the number of samples taken for the final circuit alone, and can be reduced by orders of magnitude by increasing this value in a way that actually increases total computation time. Without this $\sim100\times$ factor, the TTS advantages they claim are neutralized or inverted in favor of gate-model implementations in the majority of cases treated. We therefore do not comment on the validity or utility of the TTS definition provided, or remark on the most appropriate performance metrics with respect to runtime. Further, in sparse Max-Cut problems, we do not claim or a priori expect our quantum solver to outperform QA. Prior to posting a draft of this manuscript no annealer data was available to enable a direct comparison; Ref.~\cite{dwave_comment} introduces new results which surprisingly show the probability of finding the Max-Cut solution with our solver run on a gate-model machine is comparable to the new annealer implementation run by D-Wave. Thus, Ref.~\cite{dwave_comment} appears to independently validate and strengthen the claims of our manuscript.

\subsection*{Acknowledgments}
The authors are grateful to IBM Quantum for the provision of device access as well as access to the CPLEX classical optimization engine. We thank Andreas Bärtschi, John Golden, Stephan Eidenbenz, Elijah Pelofske, and Reuben Tate for useful discussions and the sharing of data from their previous study. We also thank Ruslan Shaydulin for sharing data and IBM device access used during early development of this work. Finally, we are grateful to all other colleagues at Q-CTRL whose technical, product engineering, and design work has supported the results presented in this paper.

\appendix
\section{Resources and run time analysis}\label{app:resources}

In previous sections, we analyzed the quality of the quantum solver compared to other solvers and previous demonstrations. Run time (cost) is another crucial metric for utility assessment. In this section, we provide details regarding the execution time and cost required for generating the data presented in this work.
In Table~\ref{table:results}, we show results from Quantinuum and D-Wave as have appeared in previous publications. Using publicly available information, we approximate the runtime and cost required to produce the results shown.

\emph{Quantum solver (IBM):} The optimization scheme we detail in the main text is designed for fast convergence. Our method batches six circuits at each optimization step in order to reduce overhead due to API and data transfer latency. The classical compute time between each two steps includes, for each of the six circuits, analysis of the data for cost extraction, the determination of the next set of circuit parameters, parameter binding and transpilation. In this work, the classical processing is performed sequentially for the six circuits and requires under 10 seconds, which totals to under two minutes throughout the full optimization loop. A straightforward parallelization of this step can effectively render the classical compute time negligible.

The smallest problems we solved (28--32 nodes Max-Cut) implemented 12 optimization steps, which totals to 72 circuits, each executed with 6,000 shots for a total of 432,000 shots throughout the full optimization process. IBM's reported repetition time for this data is $\sim4,000$ shots per second, resulting in an ideal total QPU time of 108 seconds. In practice, the measured cumulative QPU time of these jobs is about 150 seconds. Additional API latency, overheads, and queue times on the hardware provider side exist and typically match the total QPU time (about 150 seconds), leading to a total wall-clock optimization time of about six minutes. Based on published IBM pricing, execution of the full optimization loop for these problems required approximately $\$400$~USD to run.

The larger problems we consider use 16--20 optimization steps, totaling 96--120 circuits, each with 8,000--15,000 shots leading to a total of 768,000--1,800,000 shots. A full optimization loop using the maximum number of shots required 7--8 minutes of QPU time and in the worst-case 20~minutes of total wall-clock time. Cost for such executions ranges $\sim$\$1,120--\$1,280~USD.

\emph{Quantum annealers (D-Wave):} Based on Refs.~\cite{lubinski2024, pelofske2023comparing}, the dominant component of QPU time for QA is the anneal time which can be tuned between 0.5--2000~$\mu$s. In general, longer anneal times show better stability; in Ref.~\cite{pelofske2023comparing}, the best anneal times were identified to be between 1000--2000~$\mu s$. On top of the anneal time, readout time of 250~$\mu$s and a programming time of 160~$\mu$s should be added for each sample.

The QA data in Table~\ref{table:results}, for three instances of the spin-glass problem, is taken from Ref.~\cite{pelofske2024short}, where a total of $1,328,000$ samples where used for each problem. Assuming a sample time of 1000--1500~$\mu$s corresponds to 22--33 minutes of QA time.

The total optimization time includes the quantum execution time, along with the time for computing a minor embedding of the input to match the specific qubit connection structure inside the QPU, and the time to resolve solutions by mapping them back to the original (unembedded) problem. We have no information regarding the total execution time or cost for this system. Ref.~\cite{dwave_comment} from D-Wave suggests a relevant sample time of 1.6 ms, making our estimate conservative, and provides general guidance on runtime of related problems.

\emph{Trapped-ion quantum computer (Quantinuum):} Ref.~\cite{shaydulin2023qaoa}, tested QAOA executions on Quantinuum H2-1 device. The results in Table~\ref{table:results} represent a single circuit execution rather than a full iterative optimization process. We can estimate the cost of running such a circuit using available pricing information in \cite{Azure_Quan}; we are not aware whether this published pricing is indicative of commercial terms available to Quantinuum customers. A 32-node 3-regular graph has 48 edges, hence, assuming arbitrary 2-qubit rotations are considered as basis gates and all to all connectivity, such a $p=10$ QAOA circuit includes 480 2-qubit gates, 362 single-qubit gates and 64 reset/measurement operations (a requirement of higher-depth QAOA with $p>1$ as implemented in~\cite{shaydulin2023qaoa} would require a proportionate growth in circuit resources). With 1,024 shots, such a circuit consumes 1,128 H-System Quantum Credits (HQCs). With a standard subscription to H2 devices \cite{Azure_Quan}, one HQC is equivalent to $\$13.5$~USD, making the total cost of a single circuit execution $>\$15,000$~USD.

We do not know whether full convergence of the optimization process on hardware would be possible as executed or whether if successful the convergence rate would differ from the implementation we describe above. However, assuming convergence equivalent to that achieved in the tests we have performed, full loop optimization over an equivalent 72 circuits would require $\$1,080,000$~USD.

In Refs.~\cite{decross2023qubitreuse, moses2023racetrack}, the full optimization is done on the quantum computer. As reported, the optimization process for the 80-node Max-Cut problem lasted 11 hours, with an achieved approximation ratio of 91\%. Moreover, Ref.~\cite{moses2023racetrack} mentions that a single sample (shot) time for a $p=2$ QAOA of a 32-node 3-regular requires 0.97 sec while a $p=1$ QAOA on the 130-node 3-regular problem (with qubit reuse on the 32-qubit device) requires 2.83 sec. Using the information provided in Ref.~\cite{moses2023racetrack}, for the 32-node problem, 200 shots per circuit were used totaling approximately 90~minutes of QPU time. For larger and more complex problems, arriving at the optimal solution is likely to require higher-depth circuits and tens of thousands of shots.

\section{Post-processing and local solver}\label{app:classical}

\emph{Classical post-processing:} We use a na\"\i ve greedy optimization for the classical post-processing pass. Considering a bit flip to be a local move, bitstrings for which no local move improves the cost can be considered to be a local minimum of the problem. Greedy optimization corresponds to flipping single bits to improve the cost until a local minimum is reached, and no further improvement is possible. Pseudocode of the greedy optimization procedure is presented in Algorithm~\ref{alg:greedy}. There, Shuffle$(N)$ randomly shuffles the bitstring indices ${i=0,...,n-1}$, Flip$(x,i)$ modifies bitstring $x$ by flipping the $i$-th bit. We limit the number of full traversals of the bitstring to a maximum of five passes.

The motivation for this pass is that the QAOA circuit produces a nontrivial bitstring distribution with support on bitstrings in the basin of good local minima. Device noise and measurement errors can be expected to perturb the bitstrings by local moves, but even a few erroneous bit flips can significantly decrease the solution quality. Greedy optimization corrects for this. The greedy post-processing scales as $\mathcal{O}(n)$ and therefore does not introduce any additional overhead.

\emph{Local solver:} We implement a purely classical ``local solver'' that samples bitstrings uniformly at random and applies a greedy post-processing pass to generate local minima of the cost function (see Fig.~\ref{fig:error_suppression_and_postprocessing}e). Here, a local minimum is defined to be a bitstring whose cost cannot be improved by local moves, in this case single bitflips. Pseudocode for the local solver is presented in Algorithm~\ref{alg:localsolver}. After first generating a set of random bitstrings, a greedy pass (Algorithm~\ref{alg:greedy}) produces a local minimum by greedily flipping any bits that result in an improved cost (in Algorithm~\ref{alg:greedy}, Flip($\bm{x}, j)$ flips the $j$-th bit of bitstring $\bm{x}$). Importantly, the order in which the local moves are considered is randomized (as indicated by the Shuffle($n$) term in the pseudocode). The greedy pass terminates when no cost-improving local moves exist. (In practice, the greedy pass terminates after a few bitflips, however, to avoid pathological edge cases, the pass is constrained through early stopping to no more than five full traversals of the bitstring.)

\begin{algorithm}[H]
\caption{GreedyPass}\label{alg:greedy}
   \begin{algorithmic}
\Require length-$n$ bitstring $\bm{x}$, cost function $C$
\Ensure local minima bitstring
\While{$\bm{x}$ not local minima of $C$}
    \For{$j$ in Shuffle($n$)}
        \State $\bm{x}' = \text{Flip}(\bm{x}, j)$
        \If{ $C(\bm{x}') < C(\bm{x})$}
            \State $\bm{x} \leftarrow \bm{x}'$
        \EndIf
    \EndFor
\EndWhile
\ReturnX $\bm{x}$
   \end{algorithmic}
\end{algorithm}

\begin{algorithm}[H]
\caption{LocalSolver}\label{alg:localsolver}
   \begin{algorithmic}
\Require sample size $N$, cost function $C$, passes $K$
\Ensure sample of local minima
\State Initialize $\bm{x}_i$ as $N$ randomly sampled bitstrings
\For{$i$ in $1:N$}
    \State $c \gets C(\bm{x}_i)$
    \State $\bm{x}_{\text{best}} \gets \bm{x}'$
    \For{$k = 1$ to $K$}
        \State $\bm{x}' \gets$ GreedyPass($\bm{x}_i$)
        \If{$C(\bm{x}') < c$}
            \State $c \gets C(\bm{x}')$
            \State $\bm{x}_{\text{best}} \gets \bm{x}'$
        \EndIf
    \EndFor
\State $\bm{x}_{i} \gets \bm{x}_{\text{best}}$
\EndFor
\ReturnX $\{\bm{x}_i\}_{i=1, ..., N}$
   \end{algorithmic}
\end{algorithm}

Since a given bitstring may be connected by cost-reducing local moves to multiple local minima, the local solver applies the greedy pass $K$ times for each bitstring, each time using a different random seed. Since the order in which bits (spins) are traversed affects the outcome, each seed can lead to a different local minimum. The best of the $K$ minima produced is taken to be the post-processed bitstring. For this work, we set $K=5$.

We note that such a local solver may be thought of as a zero-temperature version of Simulated Annealing (SA). In SA, similar sweeps are performed, however non-greedy moves are accepted with some probability that gradually approaches zero towards the end of the annealing process. Typical SA uses 500--5000 sweeps for each initial state (typically 100--1000 such states are considered).

\section{Initial state parameterization}\label{app:initialstate}
This Appendix describes our modification to the standard QAOA initial state through parameterized biasing, the adaptive update rule that leverages intermediate results, and empirical assessment of how this enhancement improves convergence.

\subsection{Initial state definition}
The parameterized initial state corresponds to the layer of single-qubit $R_y$ rotation gates shown in Fig.~\ref{fig:ansatz}, which replaces the standard uniform superposition $\ket{s} = \ket{+}^{\otimes n}$ with a biased state that favors specific bitstrings while maintaining the ability to explore the solution space.

Beginning with a single qubit, the action of $R_y(\theta)$ on $\ket{0}$ produces:
\begin{equation}
    R_y(\theta) \ket{0} = \cos\left( \frac{\theta}{2} \right) \ket{0} + \sin\left( \frac{\theta}{2} \right) \ket{1}.
\end{equation}
The standard QAOA initial state is recovered by setting $\theta = \pi/2$ to create an equal superposition. To introduce bias toward a target bit value $b_T \in \{0, 1\}$, we instead parameterize the rotation angle as
\begin{equation}
	\label{eq:single_qubit_biasing}
    \theta = \frac{\pi}{2} - (-1)^{b_T} \delta,
\end{equation}
where $\delta \in [0, \pi/2]$ serves as the biasing parameter. This yields a measurement probability $\text{Pr}(b_T) = (1 + \sin \delta) / 2$ for the target bit. The parameter $\delta$ controls the bias strength: $\delta = 0$ recovers the uniform superposition, while $\delta = \pi/2$ produces a deterministic state $\ket{b_T}$.

For the $n$-qubit case, given a target length-$n$ bitstring $\bm{b}_T \in \{0, 1\}^n$, we construct the biased initial state as:
\begin{equation}
	\label{eq:biasing_state}
    \ket{\psi_\delta(\bm{b}_T)} = \bigotimes_{i=1}^n R_y\left( \frac{\pi}{2} - (-1)^{(b_T)_i} \delta \right) \ket{0}_i = \sum c_b \ket{\bm{b}}.
\end{equation}

Since the state is a product state, the probability of sampling any bitstring $\bm{b}$ factorizes into independent per-qubit contributions, with each qubit biased toward the corresponding target bit value by an amount controlled by $\delta$. This product state has a particularly useful property: the squared amplitude of any computational state $\ket{\bm{b}}$ depends only on the biasing angle $\delta$ and the Hamming distance $d_H(\bm{b}, \bm{b}_T)$ and is given by
\begin{equation}
    \left|c_b\right|^2 = \left( \frac{1 + \sin \delta}{2}\right)^{n} \left( \frac{1 - \sin\delta}{1 + \sin\delta} \right)^{d_H(\bm{b}, \bm{b}_T)}.
\label{eq:squaredAmplitudeComputationalState}
\end{equation}
Note that when $\delta=0$, the unbiased case, $|c_{b}|^2 = 2^{-n}$ for all $\bm{b}$.

Equation~\eqref{eq:biasing_state} defines a single-parameter family of product states where measurement probabilities from this state decrease monotonically with Hamming distance from the target bitstring.

\begin{figure}[htb]
    \centering
    \includegraphics[width=0.45\textwidth]{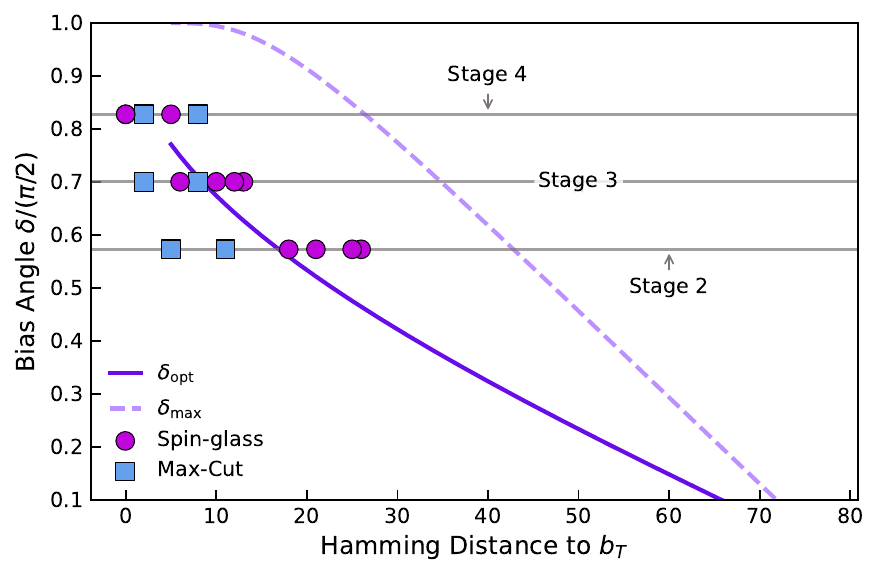}
    \caption{
    \label{fig:optimal_and_bias}
      Theoretical bounds for optimal and maximum bias angles for $n$=156. The gray horizontal lines represent the bias angle $\delta$ schedule used for all data presented in this work (the first stage in the multi-stage optimization always has $\delta=0$ recovering the standard QAOA ansatz. We include all data taken with $n=156$, which includes two Max-Cut instances and four cubic spin-glass instances. The data points represent the hamming distance between the optimal solution returned at the end of the stage and the target bitstring used for biasing the initial state for all circuits during that stage.
      }
\end{figure}

\subsection{Measurement-feedback-based initial state biasing}
When executing the solver on hardware, each circuit is sampled a number of times (according to the shot count) as part of the closed-loop variational parameter ($\bm{\beta}$, $\bm{\gamma}$) optimization procedure. Each shot results in a bitstring with an associated cost. Our algorithm leverages the intermediate bitstring cost information by introducing a multi-stage optimization procedure. Each stage consists of multiple optimization steps where the $\bm{\beta}$, $\bm{\gamma}$ variational parameters are optimized, but the initial state parameters remain fixed. Between each stage, the initial state parameters are updated. 

The initial state replaces the traditional Hadamard gates of standard QAOA with parameterized single-qubit $R_y$ gates (see Eq.~\eqref{eq:biasing_state}), with the angle defined as shown in Eq.~\eqref{eq:single_qubit_biasing}. For the first stage, $\delta$, the biasing strength parameter, is set to $0$, recovering the standard QAOA ansatz. In subsequent stages, increasing values of $\delta$ are chosen from a predetermined schedule, using the lowest-cost bitstring observed thus far as the target $b_T$. $\delta$ is chosen such that the algorithm progressively localizes the initial state superposition around promising solutions while maintaining sufficient exploration through the QAOA dynamics. Our current $\delta$ schedule was determined empirically and fixed to the same schedule for all runs presented in this work. In this section, we use the 156-qubit data collected to illustrate how the schedule chosen compares to theoretical limits.

Biasing can negatively impact our initial state if it is too strongly biased toward a bitstring $\bm{b}_T$ (assuming $\bm{b}_T$ is not an optimal solution). If the bias angle, $\delta$, is sufficiently large, the squared amplitude of an optimal solution, $\bm{b^*}$, can be less than the squared amplitude when no bias is applied. That is, for a some $d_H(\bm{b^*}, \bm{b}_T)$, there is a range $\delta \in [0, \delta_{\text{max}}]$ such that $|c_{b^*}|^2 > 2^{-n}$, and outside of this range is less than $2^{-n}$.

We can determine two important bias angles for a particular problem size from Eq.\,(\ref{eq:squaredAmplitudeComputationalState}): an optimal bias angle, $\delta_{\text{opt}}$, and a maximum bias angle, $\delta_{\text{max}}$. The optimal bias angle is found maximizing Eq.\,(\ref{eq:squaredAmplitudeComputationalState}), given by $\delta_{\text{opt}} = \sin^{-1}\left((n - 2h)/n\right)$, where $h = d_H(\bm{b}, \bm{b^*})$.
The maximum bias angle is set by finding the angle such that $|c_{b^*}|^2 = 2^{-n}$, which is equivalent to finding the non-zero angle that satisfies $(n - h) \ln (1 + \sin \delta) + h \ln (1 - \sin \delta) = 0$. 
The bias angles chosen for each stage of our algorithm are well optimized for the six 156-qubit problems instances included in this work (Figure \ref{fig:optimal_and_bias}). The bias angles for these stages are more conservative for smaller problems sizes.

The complete algorithm is presented in Algorithm~\ref{alg:biased_qaoa}, which takes as input the cost Hamiltonian $H_C$, the number of stages $S$, the number of iterations per stage $T$, and the biasing schedule $\{\delta_s\}_{s=1}^S$. Within each stage, the initial state parameterization remains fixed while the QAOA parameters are optimized using CMA-ES with CVaR as the objective function. Between stages, the initial state is updated to bias toward the best solution found, enabling efficient exploration of progressively refined search regions.

\begin{algorithm}[H]
\caption{BiasedQAOA}\label{alg:biased_qaoa}
   \begin{algorithmic}
\Require Cost function $C$, stages $S$, iterations per stage $T$, bias schedule $\{\delta_s\}_{s=1}^S$
\Ensure Optimized bitstring $\bm{x}_{\text{best}}$
\State Initialize $\bm{\theta} \gets \pi/2 \cdot \mathbf{1}_n$ \Comment{Uniform superposition}
\State Initialize $\bm{x}_{\text{best}} \gets \text{null}$
\For{stage $s = 1$ to $S$}
    \If{$s > 1$}
        \State Update $\bm{\theta}$ using Eq.~\eqref{eq:biasing_state} with $\bm{b}_T = \bm{x}_{\text{best}}$ and $\delta = \delta_s$
    \EndIf
    \For{iteration $t = 1$ to $T$}
        \State Optimize $\bm{\gamma}$, $\bm{\beta}$ using CMA-ES with CVaR objective
        \State Execute circuits and measure bitstrings
        \State Update $\bm{x}_{\text{best}}$ if better solution found
    \EndFor
\EndFor
\ReturnX $\bm{x}_{\text{best}}$
   \end{algorithmic}
\end{algorithm}

\begin{figure}[htb]
    \centering
    \vspace{-12pt}
    \includegraphics[width=0.48\textwidth]{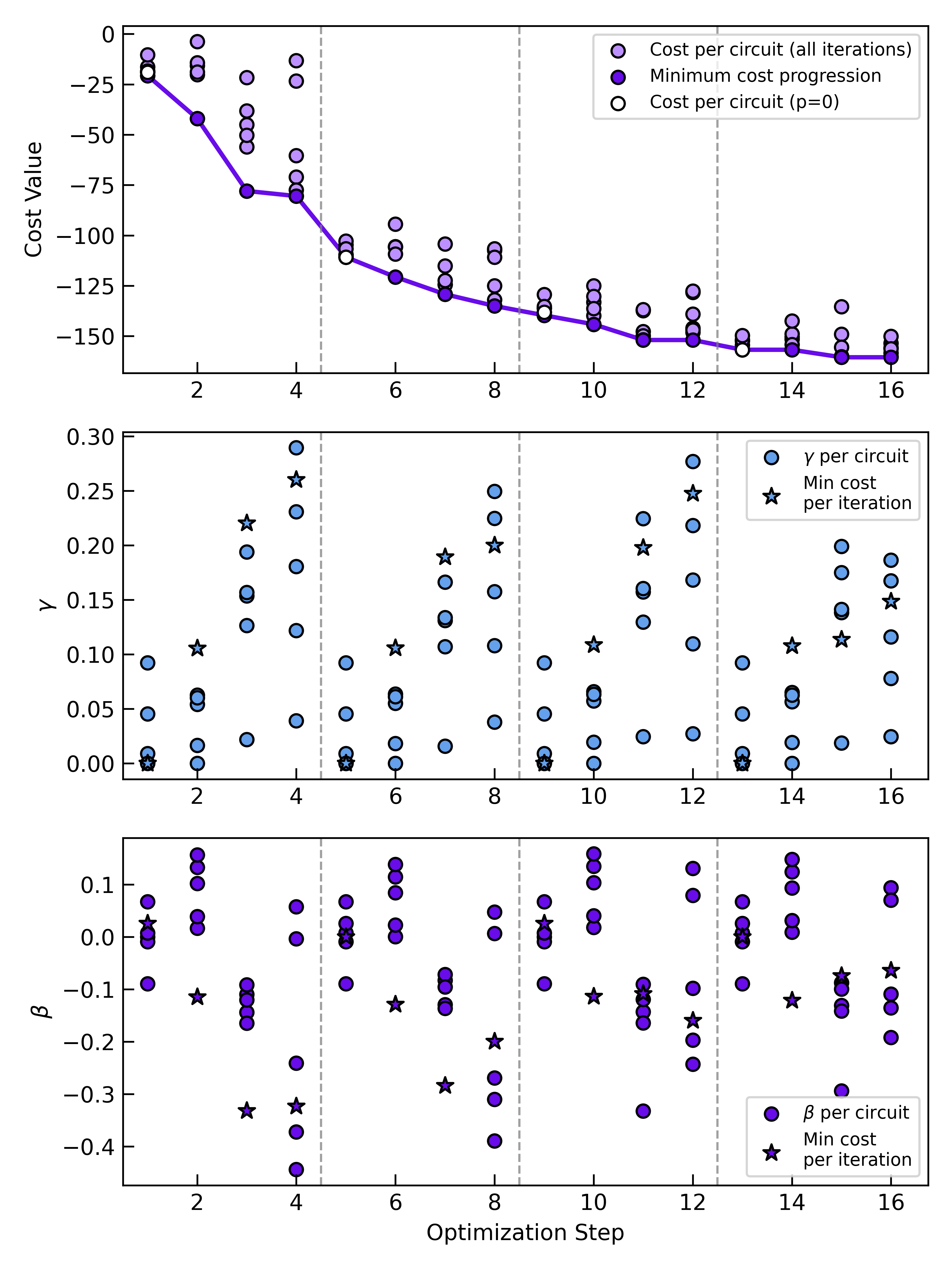}
    \vspace{-12pt}
    \caption{
    \label{fig:cvar}
      \textit{Top panel:} CVaR cost ($\alpha=0.35$) per solver iteration for spin-glass problem instance 5 (shown in Fig.~\ref{fig:maxcut_and_lanl_combined_plot}c). The dark purple data points are the minimum cost observed up to and including that optimization step, and the light purple points are the CVaR cost evaluated for each circuit during the optimization process. There are six circuit evaluations per step. The vertical lines in each panel represent each time the $\bm{\theta}$ angles were updated during the optimization. In the first step and also in the first step after each vertical line (optimization steps 1, 5, 9, and 13) we include one circuit with variational parameters $(\gamma, \beta) = (0,0)$ as a baseline for comparison (highlighted in white). The CVaR cost, which is used during the hybrid optimization loop, is evaluated for each circuit after measurement error mitigation \cite{mundada2023experimental} prior to the bitflip post-processing. \textit{Lower panels:} The values of the QAOA parameters $\gamma$ (middle) and $\beta$ (bottom) during the optimization (we use $\sigma=0.1$ for CMA-ES and $\gamma$ is restricted to positive values). While the initial QAOA parameters are generated randomly at the beginning of the optimization run, we chose to use the same random seed when generating points after the initial state updates in order to isolate and observe the role of the initial state only.
      }
\end{figure}

\subsection{Empirical evaluation}
To illustrate the impact of our biased initial state parameterization on algorithm performance, we examine the optimization performance for a representative spin-glass optimization run. Figure~\ref{fig:cvar} presents a complete optimization run for spin-glass problem instance 5, tracking the best CVaR cost evolution across all stages and iterations as well as showing the CVaR values and variational parameters for all circuits run during the optimization process.

For each of the 16 optimization steps, split into four stages, we submit six circuits to the quantum computer. Each circuit submitted within each stage has the same initial state $\bm{\theta}$ values, optimizing only the $\bm{\gamma}$ and $\bm{\beta}$ parameters in the CMA-ES loop. For the first step of each stage, we include one circuit where $\bm{\beta} = \bm{\gamma} = 0$, which is the equivalent of a $p=0$ circuit only containing the initial state followed by measurement, and for the other five circuits, the variational parameters are chosen randomly. The CMA-ES optimizer ($\sigma=0.1$) then uses the CVaR cost ($\alpha=0.35$) of each measured distribution to define the variational parameters for the next iteration. The $\bm{\theta}$ angles are all initialized to $\pi/2$ and the initial state is unchanged until the second stage. Therefore, the first four steps of the optimization are equivalent to a standard QAOA execution.

After four steps, the $\bm{\theta}$ angles are updated and a new stage begins, indicated by the vertical dashed lines in the figure. When the initial state parameters are updated, the following optimization step again includes one circuit for which the QAOA parameters $\bm{\beta} = \bm{\gamma} = 0$, equivalent to a $p=0$ version of the algorithm with the modified initial state ansatz. Accordingly, steps 1, 5, 9, and 13 in Fig.~\ref{fig:cvar} each include one $p=0$ circuit as one of the six data points used by the CMA-ES optimizer in optimization of the circuit parameters and are highlighted in Fig.~\ref{fig:cvar} with white circle markers. The initial state is unchanged within each stage; i.e., the circuits submitted to hardware in steps 1--4, 5--8, 9--12, and 13--16 have identical initial states, just different values for $\bm{\beta}, \bm{\gamma}$. Each time the initial state is updated, we again randomly choose starting values for $\bm{\beta}, \bm{\gamma}$, but we chose to use the same random seed in order to isolate the role of the initial state.

We update the initial state by using the best solution found across previous iterations to define a new nonuniform starting superposition that biases the initial state towards the target bitstring. The biasing angle $\delta$ increases for each stage. As shown in Fig.~\ref{fig:cvar}, we observe that every time the initial state is updated, there is an improvement in the CVaR cost. This may be attributed to the fact that increasing the biasing angle narrows the support of the initial state around the best-observed computational state. We also observe an improvement in the cost when the initial state is unchanged, and only the QAOA parameters are changed. This observation indicates that the QAOA circuit parameter optimization plays a role in solving the problem, beyond the initial-state parameterization.

In the same plot, we also show explicitly the values of $\bm{\beta}, \bm{\gamma}$ throughout the optimization. Each time the initial state is changed, the $\bm{\beta}, \bm{\gamma}$ angles are initialized to the same randomly generated points from the beginning of the optimization in order to isolate the performance from the initial state parameterization. We only perform this optimization for four steps before updating the initial state in order to reduce as much as possible the resources required for convergence. In Fig.~\ref{fig:cvar}, we see on average the $\bm{\beta}, \bm{\gamma}$ parameters increase in magnitude when optimized.

\begin{figure}[htb]
    \centering
    \includegraphics[width=0.48\textwidth]{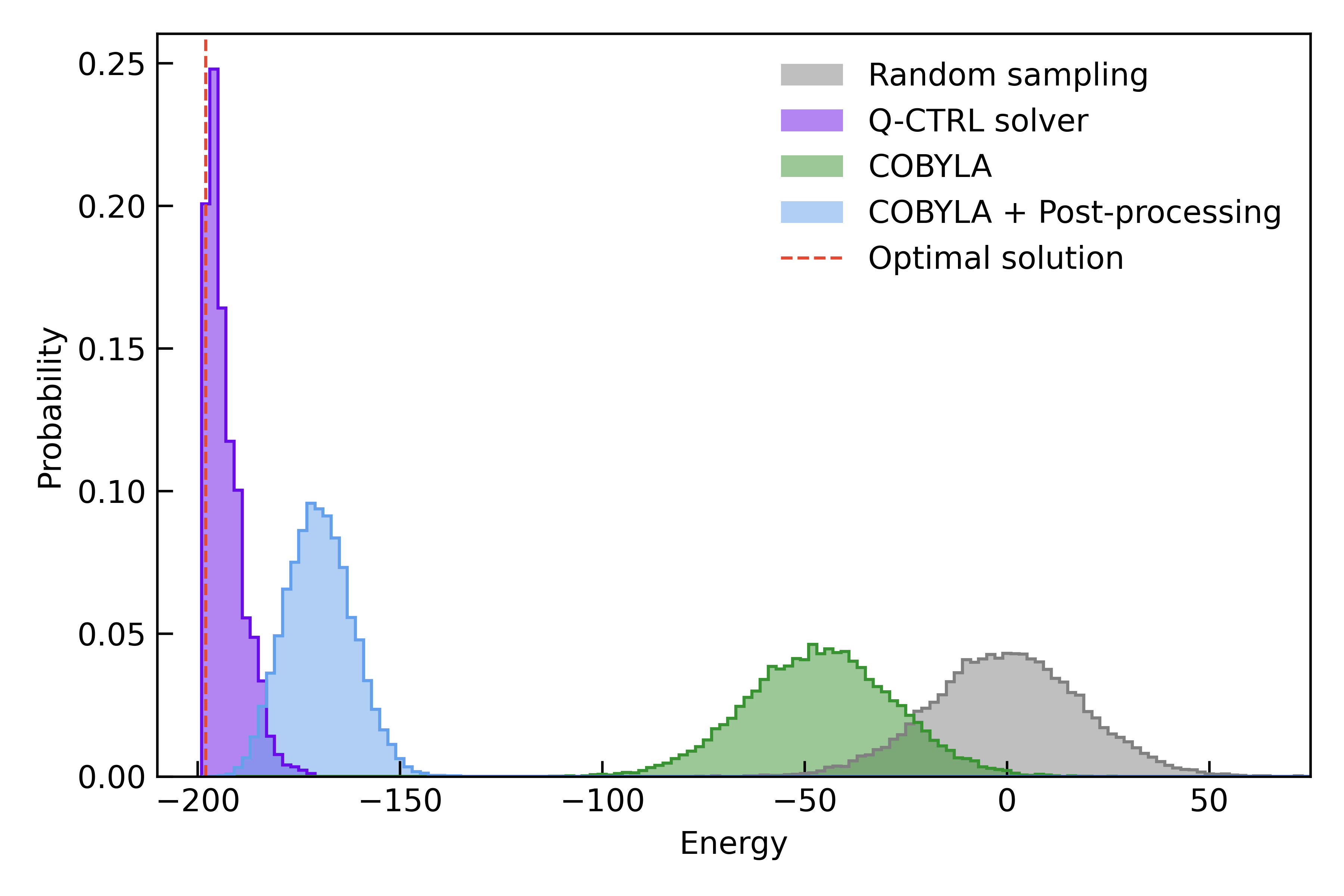}
    \caption{
    \label{fig:p0simulation}
      Initial state optimization using COBYLA. For this simulation, we use the same parameterized initial state as our modified QAOA ansatz, but remove all QAOA layers ($p=0$). Then, we use COBYLA to optimize all $R_y$ angles directly after initializing them to $\pi/2$ using the cost function from instance 5 and the same CVaR $\alpha$ (our experimental results are shown in Fig.~\ref{fig:maxcut_and_lanl_combined_plot}c and reproduced here). We run this for a maximum of 100 iterations after which the optimization is stopped. Following the optimization, we apply our postprocessing pass described in Appendix~\ref{app:classical} to the optimized distribution.
      }
\end{figure}

Lastly, we consider the possibility that the initial state parameterization alone may be responsible for the observed performance of our solver. To exclude this hypothesis, we consider a noiseless simulation of a $p=0$ circuit corresponding a simple product state with no entanglement. The circuit is taken to be simply a product of $R_y$ gates, the same as used in our initial state, but without the restriction used in the biasing, Eq.~\ref{eq:single_qubit_biasing}. Instead, the $\bm{\theta} \in \mathbb{R}^n$ angles are allowed to take on arbitrary values. These are optimized using a number of objective function evaluations chosen to match the number of circuit executions as a typical run ($\sim 100$). Here we use COBYLA rather than the CMA-ES optimizer used in the rest of this work as CMA-ES is ill-suited for this use-case given the number of angles we want to optimize (e.g., 127 or 156 in the case of the device-scale problems), and the limited circuit executions. In order to use CMA-ES with a reasonable population size per optimization step, we would have to limit the number of optimization steps to a maximum of 6, which is not sufficient for convergence. Fig.~\ref{fig:p0simulation} shows that the resulting optimized product state, with or without the greedy post-processing, is unable to match the results of our algorithm. This constitutes numerical evidence against the hypothesis that the observed performance of our solver can be attributed to the initial state parameterization alone and confirms that QAOA entanglement dynamics contribute to the observed results.

\section{Parametric compilation analysis}\label{app:compilation}

The Q-CTRL \emph{parametric compiler} effectively reduces both circuit gate counts and duration compared to other commercially available parametric compilers. Here, we compare the quality of compilation, measured in terms of the number of two-qubit entangling gates required for circuit execution. We treat two extremal problems which bound the expected envelope of circuit structures to be encountered in generic optimization problems.

\begin{figure*}[tp]
    \centering
    \includegraphics[width=0.979\textwidth]{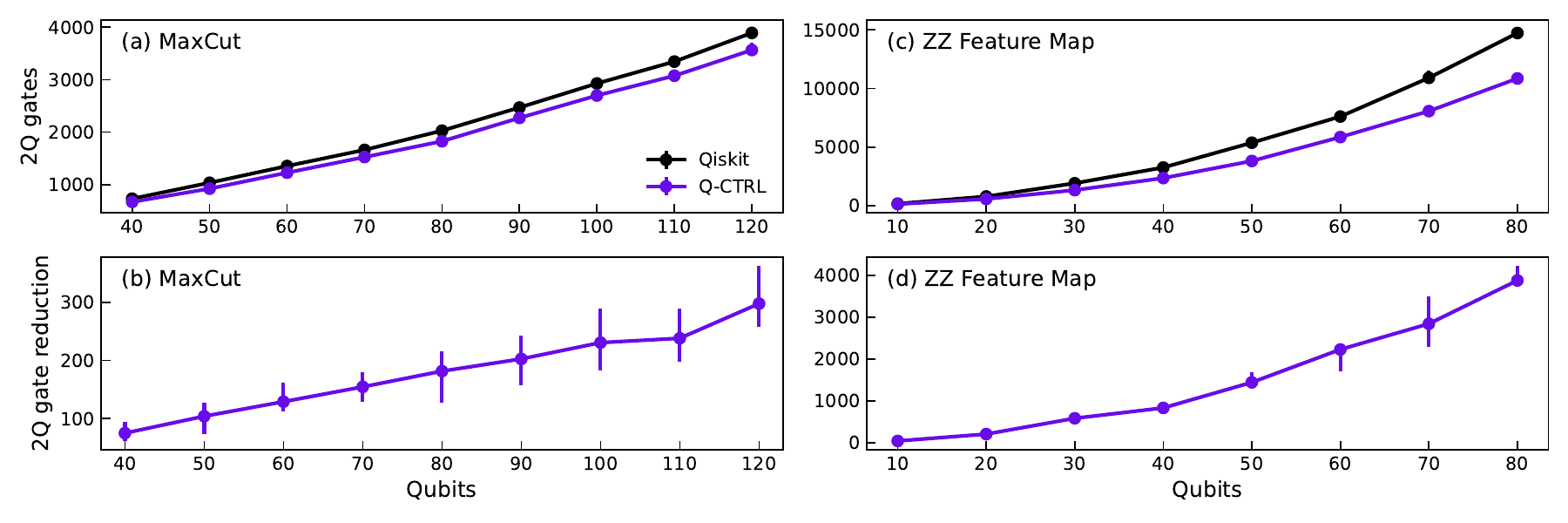}
    \caption{
    \label{fig:compiler}
    Compiler benchmarks for sparse Max-Cut and fully entangling ZZ-feature-map problems. (a-b) Compiler performance measured as the median number of 2-qubit entangling gates in compiled $p=2$ QAOA circuits over $40$ random 3-regular graphs for Max-Cut as a function of qubit number (note that for the work presented in the main text, we only use $p=1$ which uses roughly half as many 2-qubit gates). We compare our parametric compiler to Qiskit level 3 optimization. In (a), we show the total number of gates in the circuits and in (b) the difference (gain) in the 2-qubit gate count between the two compilers.
    (c-d) Similar analysis for the ZZ feature map construction \cite{ZZ_map} with fully entangled layers. Similar to (a-b) we compare our and Qiskit level compiler showing (c) the total number of 2-qubit gates and (d) the gain. This data was collected using Qiskit version 0.44.
    }
\end{figure*}

In Fig.~\ref{fig:compiler}(a-b) show $p=2$ QAOA circuits of $n$-node Max-Cut problems over 40 different random $3$-regular graphs, similar to the problems we show in the main text. We observed a $\sim 10\%$ reduction in the number of 2-qubit gates for each circuit which amounts to a reduction of $\sim 300$ gates at the maximum tested 120 nodes (qubits).

In Fig.~\ref{fig:compiler}(c-d) we consider a more generic parametric circuit structure \cite{Temme2019, ZZ_map}. In this example, we use circuits with fully entangling layers, resulting in deep circuits. Such structures are likely to appear for constrained optimization where the circuit ansatz may include non trivial mixing layers or more dense parameterization. We observed $>25\%$ reduction in the number of 2-qubit gates which amounts to a reduction of $\sim 4,000$ gates for 80-qubit circuits.

\begin{table*}[!btp]
\begin{center}
\begin{tabular*}{\textwidth}{@{\extracolsep{\fill}}ccc|cccccc|ccccc} %
\toprule
\multicolumn{14}{l}{\textbf{Max-Cut}} \\
\midrule
\multicolumn{3}{l}{\textbf{Global}} & \multicolumn{6}{|l|}{\textbf{Q-CTRL (IBM Quantum)}} & \multicolumn{5}{l}{\textbf{Local solver}}  \\
\multicolumn{3}{l}{} & \multicolumn{6}{|l|}{Full hybrid optimization} & \multicolumn{5}{l}{} \\
\midrule
Instance & Min/Max & Shots & Max & Mean & Mean AR & Count & Unique & Total shots & Max & Mean & Mean AR & Count & Unique \\
\midrule
(28, 3, 102, u)     & 0 / 40     & 6k     & 40       & 39.82            & 0.995           & 5,661             & 2                         & 432k & 40      & 38.82           & 0.971          & 3,718    & 4                        \\
(30, 3, 264, u)     & 0 / 43     & 6k     & 43       & 42.75            & 0.994           & 5,511             & 1                         & 432k & 43      & 41.43           & 0.963          & 2,810    & 2                        \\
(32, 3, 7, u)       & 0 / 46    & 6k     & 46       & 45.42            & 0.987           & 4,949             & 1                         & 432k & 46      & 43.93           & 0.955          & 2,433    & 2                        \\
(80, 3, 68, u)      & 0 / 106    & 10k    & 106      & 104.23           & 0.983           & 1,383             & 1                         & 1.20M & 106     & 101.79          & 0.960          & 1        & 1                        \\
(100, 3, 12, u)     & 0 / 135    & 10k    & 135      & 132.18           & 0.979           & 1,214             & 2                         & 1.20M & 135     & 128.86          & 0.955          & 1        & 1                        \\
(120, 3, 8, u)      & 0 / 163    & 12k    & 163      & 159.55           & 0.979           & 1,031             & 10                        & 1.15M & 163     & 155.04          & 0.951          & 2        & 2                        \\
(156, 3, 0, u) & 0 / 213     & 15k    & 213       & 210.31            & 0.987           & 3,081               & 30                        & 1.62M & 211      & 201.08           & 0.926          & 2       & 2                       \\
(156, 5, 0, u) & 0 / 324     & 15k    & 324       & 311.95            & 0.963           & 32               & 6                        & 1.53M & 322      & 309.07           & 0.954          & 1       & 1                       \\
(50, 6, 28, w) & 0 / 73.20     & 10k    & 73.20       & 71.43            & 0.976           & 810               & 4                         & 960k & 73.20      & 71.11           & 0.971          & 367      & 4                        \\
(50, 7, 27, w) & 0 / 86.25     & 10k    & 86.25       & 84.39            & 0.978           & 771               & 2                         & 960k & 86.25      & 84.04           & 0.974          & 367      & 2                        \\
(60, 5, 36, w) & 0 / 76.25     & 10k    & 76.25       & 74.01            & 0.971           & 1,013             & 5                         & 960k & 76.25      & 73.22           & 0.960          & 157      & 6                        \\
(70, 4, 75, w) & 0 / 83.75     & 10k    & 83.75       & 80.58            & 0.962           & 616               & 6                         & 960k & 83.75      & 78.96           & 0.943          & 22       & 11                       \\
(80, 4, 46, w) & 0 / 88.00     & 10k    & 88.00       & 84.01            & 0.955           & 208               & 12                        & 960k & 88.00      & 83.23           & 0.946          & 29       & 12                       \\

\midrule
\multicolumn{14}{l}{\textbf{Spin glass}} \\
\midrule
\multicolumn{3}{l}{\textbf{Global}} & \multicolumn{6}{|l|}{\textbf{Q-CTRL (IBM Quantum)}} & \multicolumn{5}{l}{\textbf{Local solver}}  \\
\multicolumn{3}{l}{} & \multicolumn{6}{|l|}{Full hybrid optimization} & \multicolumn{5}{l}{} \\
\midrule
Instance & Min/Max & Shots & Min & Mean & Mean AR & Count & Unique & Total shots & Min & Mean & Mean AR & Count & Unique \\
\midrule
0               & -200 / 192 & 15k    & -200     & -188.43          & 0.970           & 2                 & 1    & 1.44M                     & -188    & -149.98         & 0.872          & 1        & 1                        \\
3               & -198 / 184 & 15k    & -198     & -190.37          & 0.980           & 1,768             & 33     & 1.44M                  & -190    & -152.63         & 0.881          & 1        & 1                        \\
5               & -198 / 192 & 15k    & -198     & -193.15          & 0.988           & 3,045             & 129      & 1.44M                 & -188    & -151.75         & 0.881          & 4        & 4                        \\
10              & -202 / 190 & 15k    & -202     & -193.42          & 0.978           & 288               & 7           & 1.44M              & -194    & -146.14         & 0.858          & 1        & 1                        \\
11              & -180 / 196 & 15k    & -178     & -173.25          & 0.982           & 2,530             & 86         & 1.44M               & -170    & -142.98         & 0.902          & 1        & 1                        \\
69              & -190 / 212 & 15k    & -188     & -182.97          & 0.983         & 2,630             & 27          & 1.80M              & -180    & -148.52         & 0.897          & 1        & 1                        \\
0 (156Q-K)             & -242 / 242 & 15k    & -242     &      -236.99     &    0.990      & 3,254            & 471          & 1.80M              & -226    & -184.63         & 0.882          & 2        & 2                        \\
0 (156Q-B)             & -242 / 242 & 15k    & -242     & -238.14          & 0.992         & 2            & 2          & 1.80M              & -226    & -184.63         & 0.882          & 2        & 2                        \\
1 (156Q-B)             & -240 / 238 & 15k    & -240     & -238.08          & 0.996         & 6,964            & 1009          & 1.80M              &   -228  & -183.65         & 0.882          & 1        & 1                        \\
2 (156Q-B)             & -250 / 246 & 15k    & -250     & -246.43          & 0.993         & 25            & 10          & 1.80M              &   -230  & -185.63         & 0.870          & 1        & 1                        \\
\bottomrule
\end{tabular*}
\end{center}
\caption{
\label{table:results_appendix}
Additional information to Table \ref{table:results} in the main text regarding the Q-CTRL solver and the Local solver. The set of problems we show here is identical. Both the ideal Min/Max values and the ones found by each solver are shown in the main text and are presented here for completeness. \emph{Shots} represents the number times the optimal circuit, found by the quantum solver, was sampled and the number of initial configurations used by the local solver, \emph{Mean} represents the averaged objective value (max cut or min energy) across all samples (shots), \emph{Mean AR} is the approximation ratio of the mean value, \emph{Count} is the number of times the top solution, found by each solver, was sampled, \emph{Unique} is the number of unique optimal configurations each solver found and \emph{Total shots} is the total amount of shots used by the quantum solver throughout the full optimization process.
}
\end{table*}

\section{Additional data}\label{app:data}

In this section, we provide additional data about the different problem instance tested. Comprehensive information about each problem execution is presented in Table~\ref{table:results_appendix}.

There we provide for each problem instance the number of optimization steps used (each optimization step contains six circuits), and the number of shots (samples) used for each circuit. Each quantum circuit execution during the optimization process uses the same number of shots. The information regarding best solution found and likelihood always refer to the optimal circuit (minimum cost) only. Additional information on each problem instance includes: the number of times the top solution was sampled, the number of unique optimal configurations found, the mean value of the objective function across all samples and the approximation ratio (AR) of the mean.

In the main text, we present the AR with respect to the best solution identified by the solver. Another common convention evaluates the AR using the mean value over all samples. In particular, this convention is informative for small problems where the number of samples---either of the optimal circuit or throughout the full optimization process---is not negligible compared to the number of possible configurations. None of problems we consider have this property; we provide here additional information about the mean values for completeness.

\begin{figure*}[!p]
    \centering
    \includegraphics[width=1.0\textwidth]{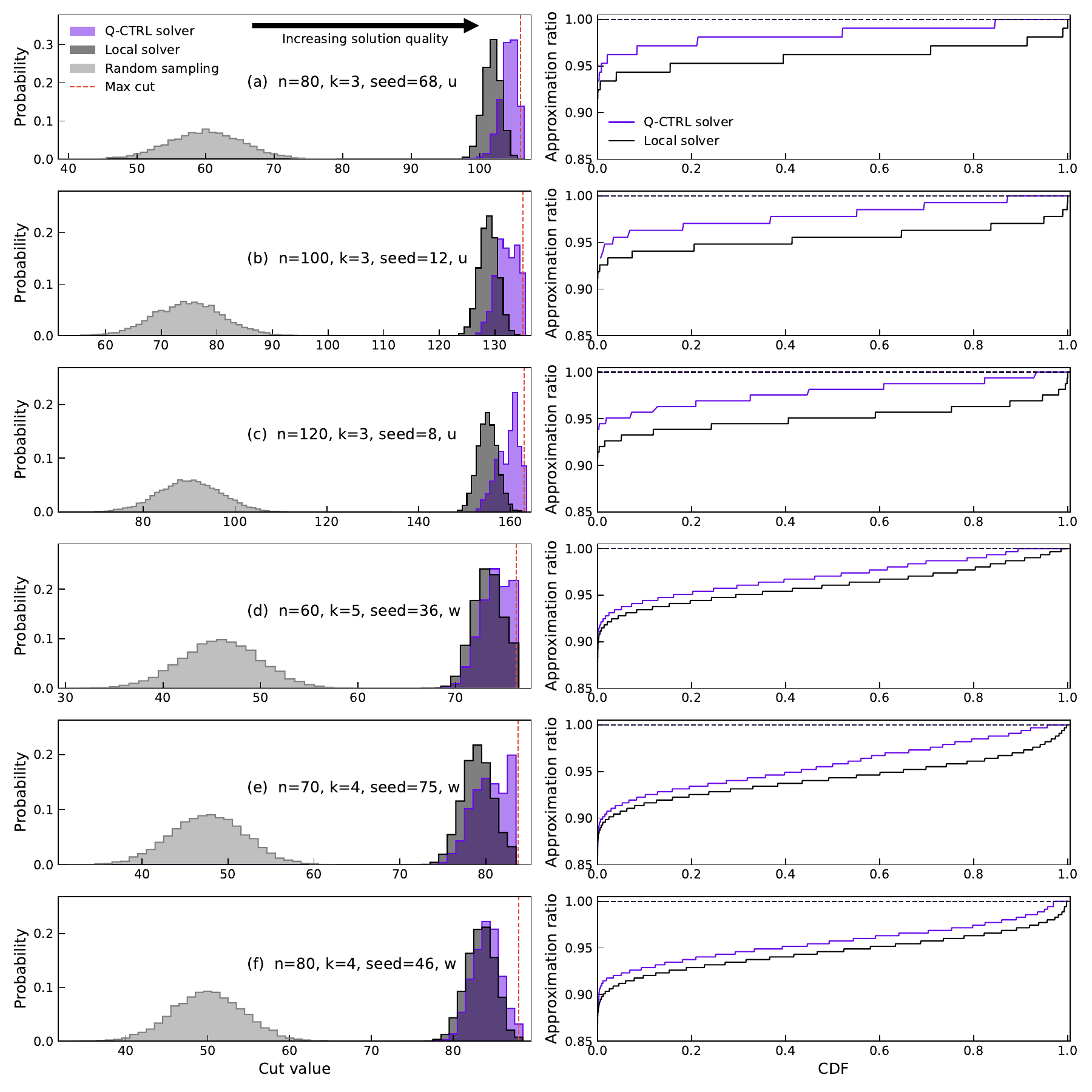}
    \caption{
    \label{fig:maxcut_cut_histogram_and_CDF_combined}
      Distribution of optimized-circuit outcomes for various Max-Cut problems.
    (Left column) The cut value distribution for $N = 6$--12k sampled configurations (see the 'Shots' column in Table~\ref{table:results_appendix}) produced by three different approaches: sampling configurations uniformly at random (Random sampling), a local solver that applies the greedy optimization of Algorithm~\ref{alg:greedy} to randomly-sampled configurations (Local solver), and the solver presented in this work. The true maximum cut value (red dashed line) is calculated using CPLEX. (Right column) The Cumulative Distribution Function (CDF) of the approximation ratio for the same data as in the adjacent panel. The vertical dashed lines indicate the maximal AR found by each solver (the correct solution corresponds to $AR=1$).
      }
\end{figure*}

\begin{figure*}[!p]
    \centering
    \includegraphics[width=1.0\textwidth]{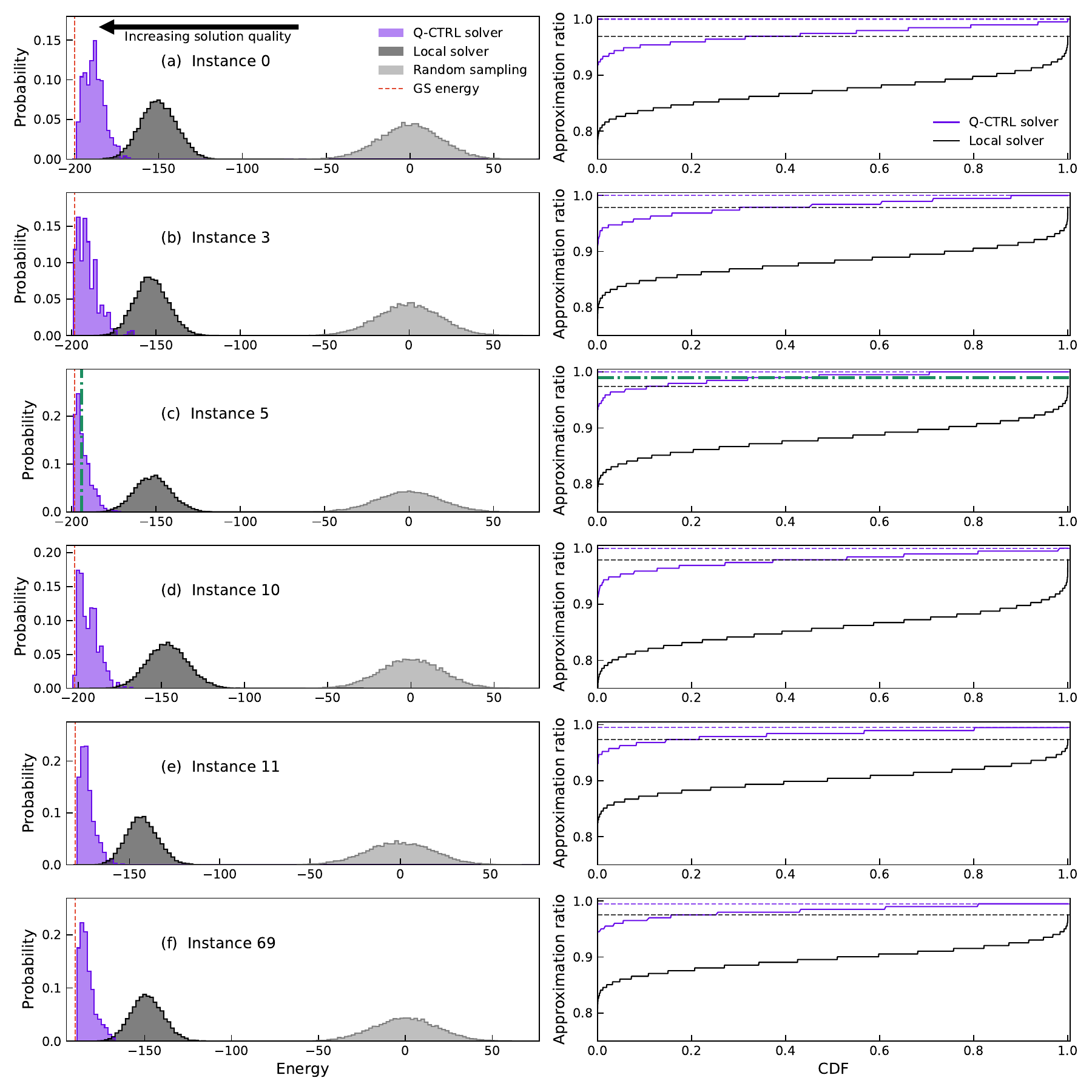}
    \caption{
    \label{fig:combined_greedy_optim_approx_ratio}
      Distribution of optimized-circuit outcomes for the spin-glass problem instances on 127 qubit IBM devices.
    (Left column) The energy distribution for 15,000 sampled configurations produced by three different approaches: sampling configurations uniformly at random (Random sampling), a local solver that applies the greedy optimization of Algorithm~\ref{alg:greedy} to randomly-sampled configurations (Local solver), and the Q-CTRL solver. The true ground-state energy (red dashed line) is calculated using CPLEX. (Right column) The Cumulative Distribution Function (CDF) of the approximation ratio (AR) for the same data as in the adjacent panel. The vertical dashed lines indicate the maximal AR found by each solver (the correct solution corresponds to $AR=1$).
      }
\end{figure*}

\begin{figure*}[!p]
    \centering
    \includegraphics[width=1.0\textwidth]{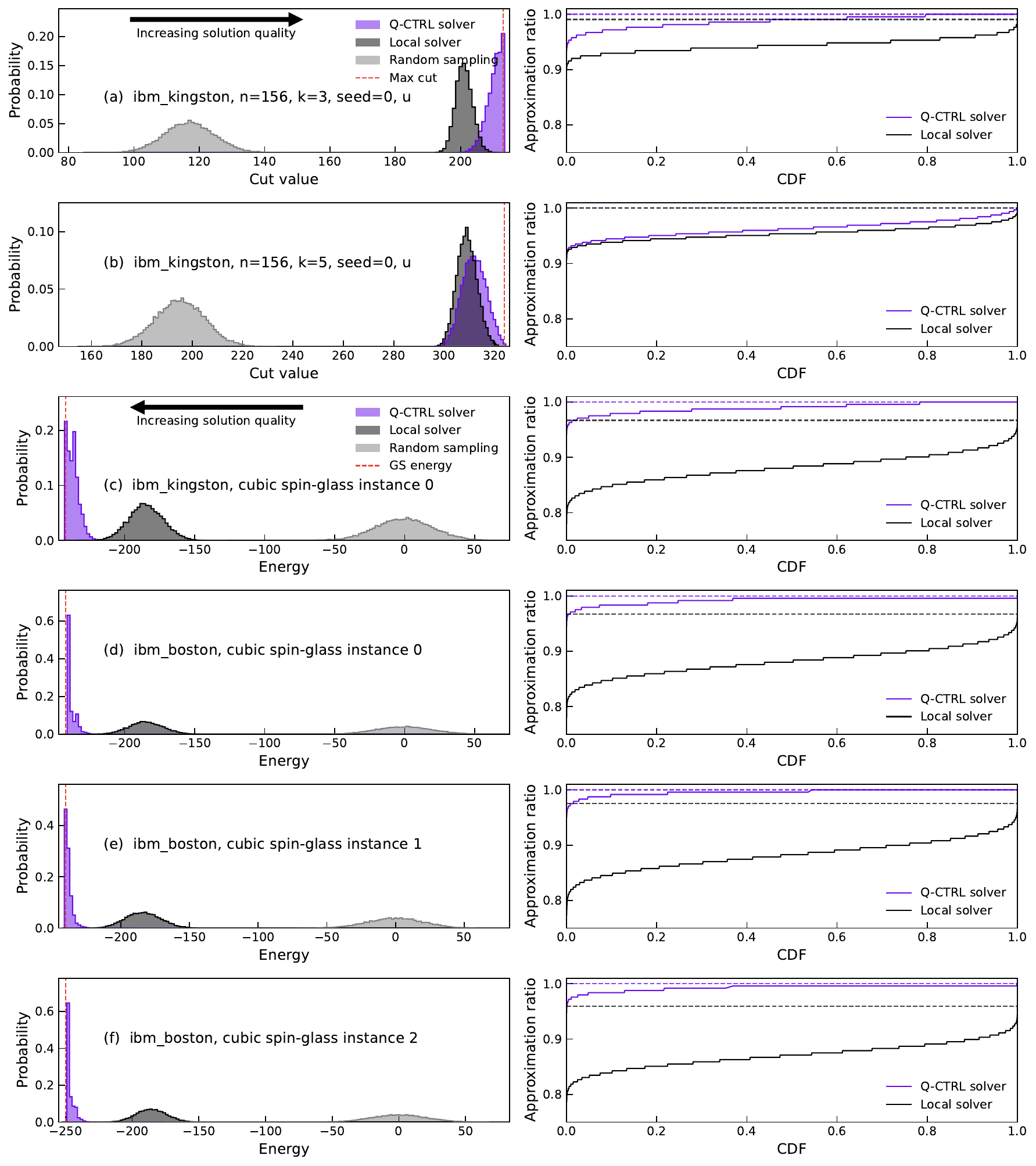}
    \caption{
    \label{fig:combined_greedy_optim_approx_ratio_156}
      Distribution of optimized-circuit outcomes for the Max-Cut and spin-glass problem instances on 156 qubit devices.
    (Left column) The energy distribution for 15,000 sampled configurations produced by three different approaches: sampling configurations uniformly at random (Random sampling), a local solver that applies the greedy optimization of Algorithm~\ref{alg:greedy} to randomly-sampled configurations (Local solver), and the Q-CTRL solver. The optimal cut value for Max-Cut and true ground-state energy for the spin-glass (red dashed line) is calculated using CPLEX. (Right column) The Cumulative Distribution Function (CDF) of the approximation ratio (AR) for the same data as in the adjacent panel. The vertical dashed lines indicate the maximal AR found by each solver (the correct solution corresponds to $AR=1$).
      }
\end{figure*}

In Figs.~\ref{fig:maxcut_cut_histogram_and_CDF_combined}--\ref{fig:combined_greedy_optim_approx_ratio}, we present further information for selected problem instance from Table~\ref{table:results}. Fig.~\ref{fig:maxcut_cut_histogram_and_CDF_combined} six selected Max-Cut instances and Fig.~\ref{fig:combined_greedy_optim_approx_ratio} shows the six 127-qubit spin-glass problem instances studied. Fig.~\ref{fig:combined_greedy_optim_approx_ratio_156} shows all 156-qubit problems studied (both Max-Cut and cubic spin-glass instances). 

In all, the left column shows the objective function (ground state energy or maximum cut value) distribution for $N$ (given for each problem in Table~\ref{table:results_appendix}) sampled configurations produced by three different approaches: sampling configurations uniformly at random (Random sampling), a local solver that applies the greedy optimization of Algorithm~\ref{alg:greedy} to randomly-sampled configurations (Local solver), and the Q-CTRL solver. The true ground-state energy (red dashed line) is calculated using CPLEX. The right column shows the Cumulative Distribution Function (CDF) of the approximation ratio for the same data as in the adjacent panel. The vertical dashed lines indicate the maximal AR found by each solver. The x-axis value indicates which fraction of the sampled configurations (out of $N$) has an AR equal or larger than the value on the y-axis.

\clearpage
\bibliography{refs}

\end{document}